\documentclass[aps,nofootinbib,superscriptaddress,twocolumn,prd,10pt]{revtex4-2}

\usepackage{amsmath,amssymb}
\usepackage{hyperref}
\usepackage{braket}
\usepackage{booktabs}
\usepackage{subfigure}
\usepackage{float}
\usepackage[dvipsnames]{xcolor}
\usepackage{mathrsfs}
\usepackage{comment}
\usepackage{bm}
\usepackage[normalem]{ulem}

\hypersetup{
    colorlinks=true,
    linkcolor=red,
    citecolor=blue,
}

\newcommand{\mav}[1]{$\langle\mu\rangle$}

\usepackage{natbib}
\usepackage[utf8]{inputenc}
\usepackage{setspace}
\usepackage[margin=1in]{geometry}
\usepackage{graphics}      % standard graphics specifications
\usepackage[pdftex]{graphicx}      % alternative graphics specifications
\usepackage{xcolor}
\usepackage{xspace}

\newcommand{\fnl}{\ensuremath{f_\mathrm{NL}}}

\newcommand{\planck}{\textit{Planck}\xspace}
\newcommand{\litebird}{\textit{LiteBIRD}\xspace}
\newcommand{\fsky}{\ensuremath{f_\mathrm{sky}}}
\newcommand{\lknee}{\ensuremath{\ell_\mathrm{knee}}}

\newcommand{\lmax}{\ensuremath{\ell_\mathrm{max}}}
\newcommand{\lmin}{\ensuremath{\ell_\mathrm{min}}}

\begin{document}

\title{CMB-S4: Forecasting Constraints on \fnl\ Through $\mu$-distortion Anisotropy}
\author{David Zegeye}
\email{dzegeye@uchicago.edu}
\affiliation{Department of Astronomy \& Astrophysics, The University of Chicago, Chicago, IL 60637, USA}
\affiliation{Kavli Institute for Cosmological Physics, The University of Chicago, Chicago, IL 60637, USA}

\author{Federico Bianchini}
\affiliation{Kavli Institute for Particle Astrophysics and Cosmology, Stanford University, 452 Lomita Mall, Stanford, California 94305, USA}

\author{J.~Richard Bond}
\affiliation{Canadian Institute for Theoretical Astrophysics, 60 St. George Street, University of Toronto, Toronto, ON, M5S 3H8, Canada}

\author{Jens Chluba}
\affiliation{Jodrell Bank Centre for Astrophysics, Alan Turing Building, University of Manchester, Manchester M13 9PL}

\author{Thomas Crawford}
\affiliation{Department of Astronomy \& Astrophysics, The University of Chicago, Chicago, IL 60637, USA}
\affiliation{Kavli Institute for Cosmological Physics, The University of Chicago, Chicago, IL 60637, USA}

\author{Giulio Fabbian}
\affiliation{Center for Computational Astrophysics, Flatiron Institute, 162 5th Avenue, New York, NY 10010, USA}
\affiliation{School of Physics and Astronomy, Cardiff University, The Parade, Cardiff, CF24 3AA, UK}

\author{Vera Gluscevic}
\affiliation{Department of Physics and Astronomy, University of Southern California, Los Angeles, California 90089, USA}

\author{Daniel Grin}
\affiliation{Department of Physics and Astronomy, Haverford College, 370 Lancaster Avenue, Haverford, PA 19041, USA}

\author{J.~Colin Hill}
\affiliation{Department of Physics, Columbia University, New York, New York 10027, USA}

\author{P. Daniel Meerburg}
\affiliation{Van Swinderen Institute for Particle Physics and Gravity, University of Groningen, Nijenborgh 4, 9747 AG Groningen, The Netherlands}

\author{Giorgio Orlando	}
\affiliation{Van Swinderen Institute for Particle Physics and Gravity, University of Groningen, Nijenborgh 4, 9747 AG Groningen, The Netherlands}

\author{Bruce Partridge}
\affiliation{Department of Physics and Astronomy, Haverford College, 370 Lancaster Avenue, Haverford, PA 19041, USA}

\author{Christian L.~Reichardt}
\affiliation{School of Physics, University of Melbourne, Parkville, VIC 3010, Australia}

\author{Mathieu Remazeilles}
\affiliation{Instituto de Fisica de Cantabria (CSIC-UC), Avda. los Castros s/n, 39005 Santander, Spain}

\author{Douglas Scott}
\affiliation{Department of Physics and Astronomy, University of British Columbia, 6225 Agricultural Road, Vancouver, BC V6T 1Z1, Canada}

\author{Edward J.~Wollack}
\affiliation{NASA/Goddard Space Flight Center, Greenbelt, MD 20771, USA}

\author{The CMB-S4 Collaboration}

\date{\today}

\begin{abstract}
Diffusion damping of the cosmic microwave background (CMB) power spectrum results from imperfect photon-baryon coupling in the pre-recombination plasma. 
At redshift $5 \times 10^4 < z < 2 \times 10^6$, the plasma acquires an effective chemical potential, and energy injections from acoustic damping in this era create $\mu$-type spectral distortions of the CMB. 
These $\mu$ distortions trace the underlying photon density fluctuations, probing the primordial power spectrum in short-wavelength modes $k_\mathrm{S}$ over the range $50 \  \mathrm{Mpc}^{-1} \lesssim k \lesssim 10^4 \ \mathrm{Mpc}^{-1}$. Small-scale power modulated by long-wavelength modes $k_\mathrm{L}$ from squeezed-limit non-Gaussianities introduces cross-correlations between CMB temperature anisotropies and $\mu$ distortions. 
Under single-field inflation models, $\mu \times T$ correlations measured from an observer in an inertial frame should vanish up to a factor of $(k_\mathrm{L}/k_\mathrm{S})^2 \ll 1$. Thus, any measurable correlation rules out single-field inflation models. 
We forecast how well the next-generation ground-based CMB experiment CMB-S4 will be able to constrain primordial squeezed-limit non-Gaussianity, parameterized by \fnl, using measurements of $C_{\ell}^{\mu T}$ as well as $C_{\ell}^{\mu E}$ from CMB $E$ modes. Using current experimental specifications and foreground modeling, we expect $\sigma(\fnl) \lesssim 1000$. This is roughly four times better than the current limit on \fnl\ using $\mu \times T$ and $\mu \times E$ correlations from \planck and is comparable to what is achievable with \litebird, demonstrating the power of the CMB-S4 experiment. This measurement is at an effective scale of $k \simeq 740 \  \text{Mpc}^{-1}$ and is thus highly complementary to measurements at larger scales from primary CMB and large-scale structure.
\end{abstract}

\maketitle

\section{Introduction}

The paradigm of cosmic inflation presents arguably the
most compelling and plausible scenario for the earliest
moments of the existence of our Universe and provides a 
mechanism for creating the density perturbations that have
grown under the influence of gravity to form all the structure
we see in the Universe today. 
(For an overview of the theoretical background and observational evidence for inflation, see, e.g., the review by Ellis \& Wands in Chapter 23 in Ref.~\cite{ParticleDataGroup:2022pth}.)
If a period of inflation did
occur, observations of the cosmic microwave background (CMB)
and tracers of the density field can give us clues about 
the nature of the fields involved in inflation and their 
dynamics. By extension, these observations can probe physics at the energy 
scale of the inflationary potential, far beyond the energy
reach of any terrestrial experiment. The CMB has most famously been used to constrain inflation models through measurements of the power spectrum of density fluctuations and constraints on the imprint of inflationary 
gravitational-wave background on the polarization of the 
CMB (e.g., \cite{BICEP:2021xfz}), but CMB observations can also shed light on inflationary
dynamics through searches for signatures of primordial 
non-Gaussianity (PNG, e.g., \cite{Silverstein:2003hf,Arkani-Hamed:2003juy,Alishahiha:2004eh,Chen:2006nt,Cheung:2007st,Senatore:2009gt,Chen:2009zp,Tolley:2009fg, Cremonini:2010ua, Achucarro:2010da,Baumann:2011nk,Barnaby:2011pe,Arkani-Hamed:2015bza}). 

The most well-studied form of PNG is the so-called  squeezed-limit bispectrum (or Fourier-domain three-point function), which describes a configuration of
triangles in Fourier space in which two of the $k$ modes
have a much larger value than the third. This corresponds
to a real-space configuration in which a long-wavelength
mode modulates the amplitude of small-scale fluctuation power.
Mathematically, this is usually expressed in terms of the 
relationship of the 
bispectrum of curvature perturbations to the curvature
power spectrum (e.g., \cite{Smith:2011if}):
\begin{equation}
    B_\zeta(k_1,k_2,k_3)_{k_3 \ll k_1 \simeq k_2} = \frac{12}{5} \fnl P_\zeta(k_1) P_\zeta(k_3),
\label{eqn:bisp}
\end{equation}
where \fnl\ parameterizes the
amplitude of local non-Gaussianity and is defined explicitly through the first-order expression for real-space local PNG:
\begin{equation}
    \zeta(x) = \zeta_g + \frac{3}{5}\fnl (\zeta^2_g - \langle \zeta^2_g \rangle ),
\end{equation}
where $\zeta_g$ is a random Gaussian field.
One of the reasons the squeezed-limit
configuration is so well-studied is that models of 
inflation with 
a single scalar field whose kinetic energy is always much
less than its potential energy (``single-field, slow-roll
inflation'') produce vanishingly small amounts of 
PNG of the squeezed-limit type \cite{Creminelli:2004yq}.
A detection of this type of PNG at levels of 
$\fnl \gtrsim 0.01$ would thus rule out 
large classes of inflation models, including many of the
most popular and viable models (e.g., \cite{CMB-S4:2016ple}).

The best current limits on the value of \fnl\ 
in the squeezed limit come 
from analyses of the primary CMB from the 
\planck satellite \cite{Planck:2019kim}: 
$\fnl = -0.9 \pm 5.1$ (68\% CL). 
Even cosmic-variance-limited maps of the 
CMB temperature and polarization out to $\ell = 4000$ would only improve
these limits by roughly a factor of five \cite{Kalaja:2020mkq}. Interest is thus 
high in other methods of determining \fnl. From a pure
mode-counting perspective, there is much more information
in the distribution of matter in the local and 
moderate-redshift Universe, but the 
non-Gaussianity caused by non-linear growth complicates
bispectrum measurements of the galaxy distribution
significantly. Previous work has shown that
squeezed-limit PNG causes a unique scale-dependent bias in the 
galaxy distribution \cite{Dalal:2007cu}, and this signature
is a target of upcoming galaxy surveys (e.g., \cite{Annis:2022xgg}), as is a direct
measurement of the matter bispectrum supplanted with careful 
modeling (e.g., \cite{Dore:2014cca}).

Another promising avenue for detecting or constraining
local PNG involves distortions to the blackbody spectrum
of the CMB \citep{Sunyaev:1970er, Burigana:1991eub, Hu:1992dc, Chluba:2011hw}. Injections of energy into the primordial plasma 
will distort the spectrum of the CMB, and a known source of
energy release in the early Universe is the dissipation of 
small-scale acoustic waves \citep{Sunyaev:1970plh, Daly:1991uob, Hu:1994bz, Chluba:2012gq} from photon diffusion or Silk damping 
\cite{Silk:1967kq}. 
The authors of Refs.~\cite{Pajer:2012vz, Ganc:2012ae} pointed out that
a modulation of small-scale power by long-wavelength
modes that underlie squeezed-limit PNG would result in
\textit{anisotropy} in the $\mu$ distortion and that a 
cosmic-variance-limited measurement of the $\mu$-distortion $-$ temperature
cross-spectrum ($\mu \times T$) could limit local-type PNG to 
$\fnl \lesssim 10^{-3}$.
Furthermore, because the $\mu$-distortion signal is created by perturbations with wavenumbers $k\simeq 50-10^4\,{\rm Mpc}^{-1}$, scales that are inaccessible by other cosmological probes \citep{Chluba:2012gq, Khatri:2012rt, Chluba:2012we}, constraints on \fnl\ from $\mu \times T$ are highly complementary to constraints on large-scale PNG from primary CMB or even large-scale structure. 

In Refs.~\cite{Remazeilles:2018kqd,Remazeilles:2021adt}, the authors performed realistic forecasting
for the limits on \fnl\ using $\mu$-distortion anisotropy
measurements from proposed space missions, including the effects
of foreground contamination. 
Similar methods have also been used to derive limits on PNG from \planck \cite{Khatri:2015tla,Rotti:2022ifo} and COBE/FIRAS \cite{Bianchini:2022dqh}, with the most stringent limit, $\fnl \lesssim 6800$ (95\% C.L.), coming from Ref.~\cite{Rotti:2022ifo}.

In this work we consider
the potential of measuring $\mu$-distortion anisotropy
from the ground. Ground-based measurements have not
been considered in previous works, mainly because the
calibration and stability requirements for measuring
the global $\mu$ signal are so stringent that it has
been assumed that only in the exquisitely stable
environment of space, with no intervening atmosphere,
would such a measurement be possible. Following Ref.~\cite{Ganc:2012ae}, we point out in
this work that measuring the $\mu$-distortion
{\it anisotropy} is a fundamentally different task.
Since anisotropy measurements can be made with a 
differencing radiometer, most of the stability and
calibration requirements are converted to requirements
on instantaneous sensitivity.

In this work we focus on the potential of the upcoming 
CMB-S4 experiment \cite{Abazajian:2019eic} to measure $\mu$-distortion 
anisotropy---specifically, the correlation between
$\mu$ and CMB temperature and polarization anisotropies. We calculate
the constraints on \fnl\ from CMB-S4 
$\mu \times T$ and $\mu \times E$ measurements, first considering 
raw sensitivity alone then adding the effects 
of atmospheric contamination and foregrounds. 

The paper is organized as follows. We discuss the
theoretical background in Section~\ref{sec:theory}; 
we describe the methods we use for forecasting in 
Section~\ref{sec:methods}; we present the survey
configuration for CMB-S4 in Section~\ref{sec:surveys};
we present our results in Section~\ref{sec:results};
and we conclude in Section~\ref{sec:discussion}.

\section{Background}
\label{sec:theory}

In this section, we review some of the key theoretical concepts in calculating and forecasting the $\mu \times T$  and $\mu \times E$ correlation arising from squeezed-limit non-Gaussianity. 

\subsection{Power Spectrum and Bispectrum}
\label{sec:powspec}

In the limit of purely Gaussian-distributed primordial curvature perturbations $\zeta({\vec{k}})$, the auto-correlation of these perturbations is given by
\begin{equation}
\left\langle\zeta_{\vec{k}_{1}} \zeta_{\vec{k}_{2}}\right\rangle=(2 \pi)^{3} \delta\left(\vec{k}_{1}+\vec{k}_{2}\right) P_{\zeta}\left({k}_{1}\right).
\end{equation}
The power spectrum $P_{\zeta}(k)$ of these perturbations from single-field inflation models is given by
\begin{equation}
    P_{\zeta}(k)=\frac{2 \pi^{2}}{k^{3}} \Delta^2_{\zeta}(k), \quad \Delta^2_{\zeta}(k)\equiv A_{s} \left(\frac{k}{k_\mathrm{p}}\right)^{n_\mathrm{s} - 1}.
\end{equation}
Here we use best-fit numbers from the \planck 2018 data release \cite{Planck:2018vyg}: $A_\mathrm{s} = 2.1\times10^{-9}$; and $n_\mathrm{s}=0.965$, reported for a pivot scale $k_\mathrm{p} = 0.05\ \textrm{Mpc}^{-1} $. 

In the squeezed limit, where the global power spectrum of two modes on short scales $k_\mathrm{S}$ are modulated by a long-wavelength mode $k_\mathrm{L}$, the long-wavelength modes generate a local non-Gaussianity and the power spectrum becomes position-dependent
\begin{equation}
    P_\zeta\left(k_\mathrm{S}, x\right)=P_\zeta\left(k_\mathrm{S}\right)\left[1+\frac{d \ln P_\zeta\left(k_\mathrm{S}\right)}{d \zeta_\mathrm{L}} \zeta_\mathrm{L}(x)\right].
\label{eq:modulate}
\end{equation}

One common measure of the level of non-Gaussianity in the curvature distribution is the Fourier-domain statistic known as the curvature bispectrum $B_\zeta(k_1,k_2,k_3)$. Analogous to the power spectrum and the auto-correlation, the bispectrum is defined through its relationship to the three-point correlation function: 
\begin{equation}
     \left\langle\zeta_{\vec{k}_{1}} \zeta_{\vec{k}_{2}}  \zeta_{\vec{k}_{3}} \right\rangle=(2 \pi)^{3} \delta\left(\vec{k}_{1}+\vec{k}_{2} + \vec{k}_{3}\right) B_\zeta(k_1,k_2,k_3).
\end{equation}
In the squeezed limit, ${k_1 / k_3 \rightarrow 0}$, the bispectrum can be expressed as
in Eq.~(\ref{eqn:bisp}), 
with \fnl\ parameterizing the amount of local non-Gaussianity coupling the power spectrum at long and short wavelengths. 
In the squeezed limit, \fnl\ obeys a consistency relationship with the primordial power spectrum \cite{Maldacena:2002vr}:
\begin{equation}
    \lim _{k_\mathrm{L} / k_\mathrm{S} \rightarrow 0} \frac{12}{5} \fnl \left(k_\mathrm{L}, k_\mathrm{S}, k_\mathrm{S}\right)=-\frac{d \ln \Delta_{\zeta}^{2}\left(k_\mathrm{S}\right)}{d \ln k_\mathrm{S}} .
\end{equation}
For slow-roll single-field inflation, the consistency relationship becomes

\begin{equation}
    \lim _{k_\mathrm{L} / k_\mathrm{S} \rightarrow 0} \frac{12}{5} \fnl \left(k_\mathrm{L}, k_\mathrm{S}, k_\mathrm{S}\right)=1-n_\mathrm{s} .
\end{equation}

For the specific observable $\mu \times T$, however, the type of PNG produced in single-field inflation results in a vanishingly small signal, far below what is predicted by the consistency relation \cite{Pajer:2012vz,Cabass:2018jgj}. This means any measurable signal of $\mu \times T$ would rule out single-field inflation models. We review the reasoning behind this result in Section~\ref{sec:sfi}.

\subsection{$\mu$ distortions}

In the early Universe, energy injected into the plasma will efficiently thermalize through double Compton scattering and bremsstrahlung, producing a blackbody distribution with a new temperature.
These two processes are highly efficient until a redshift $z_i \simeq 2\times10^6$, at which point the Universe has expanded enough that the density necessary for double Compton and bremsstrahlung to frequently occur becomes too low \cite{Hu:1992dc}. This leaves elastic Compton scattering, which conserves photon number, as the primary method for the photon-baryon bath to reach thermal equilibrium. The inefficient thermalization introduces a chemical potential, where the mixing of different blackbody spectra at different temperatures produces a Bose-Einstein rather than Planckian distribution \cite{Zeldovich:1969ff}. At small chemical potentials, the Bose-Einstein distribution can be approximated as a distorted blackbody spectrum in which 
\begin{equation}
\label{spectrum}
n(\nu) = \left[e^{h \nu /\left(k_\mathrm{B} T\right)}-1\right]^{-1} \rightarrow\left[e^{h \nu /\left(k_\mathrm{B} T\right)+\mu}-1\right]^{-1},
\end{equation}
where $h$ is Planck's constant and $k_\mathrm{B}$ is Boltzmann's constant.
This distortion of the CMB spectrum from a typical blackbody is known as a $\mu$-type distortion. 
It occurs until a redshift $z_\mathrm{f} \simeq 5\times10^4$, where even thermalization through single Compton becomes inefficient and distortions of the $y$-type start being produced \cite{Chluba:2016bvg}.

$\mu$ distortions can be generated from a variety of mechanisms. The primary contribution we will consider in this work is from diffusion damping of small-scale power. 
The $\mu$ distortions produced by diffusion damping can be related to the primordial curvature perturbations via
\begin{equation}
\mu(x) = \int_{\vec{k}_1}\int_{\vec{k}_2}\zeta(\vec{k}_1)\zeta(\vec{k}_2)   W(\vec{k}_{1},\vec{k}_{2})  e^{i (\vec{k}_{1} + \vec{k}_2) \cdot \vec{x}},
\end{equation}
where $W(\vec{k}_{1},\vec{k}_{2})$, the window function, captures the weighted amount of dissipated modes with wavenumber $k$ \cite{Chluba:2016aln, Cabass:2018jgj, Zegeye:2021yml}.  To relate the amount of $\mu$ distortions to the primordial power spectrum, we take the ensemble average of $\mu(x)$ and  ${k}_1 = {k}_2$
\begin{equation}
 \langle\mu(x) \rangle \simeq \frac{1}{2 \pi^2}\int d k P_\zeta(k) k^2 W(k).
\end{equation}
Assuming that $\mu$ distortions are generated from sub-horizon modes at the time of dissipation, the window function is
\begin{eqnarray}
    W(k)&\approx&-4.54 k^{2}  \int_{0}^{\infty}{d}z  \frac{d k_\mathrm{D}^{-2}}{dz}  \mathcal{J}_{\mu}(z)\mathrm{e}^{-\frac{2k^{2}}{k_\mathrm{D}^{2}(z)}},
\end{eqnarray}
where $k_\mathrm{D}$ is the damping scale for energy injection from diffusion damping,
\begin{equation}
k_\mathrm{D}(z) \approx 4.1\times10^{-6}(1+z)^{3/2} \textrm{Mpc}^{-1},
\end{equation}
and $\mathcal{J}_\mu$ is the time window function for $\mu$ distortions \cite{Chluba:2016aln, Cabass:2018jgj, Zegeye:2021yml}. The time window function is well approximated by an analytical Green's function given in Ref.~\cite{Chluba:2013wsa}
\begin{eqnarray}
\label{eq:derive}
    \mathcal{J}_{\mu}\left(z\right) &\approx&  \left[1-\exp \left(-\left[\frac{1+z}{5.8 \times 10^{4}}\right]^{1.88}\right)\right] \times \nonumber\\
    &&\qquad
    e^{-(z/z_i)^{5/2} },
\end{eqnarray}
where $z_i \simeq 2 \times 10^6$ is defined above as the beginning of the $\mu$-distortion era.
For single-field, slow-roll inflation, the average amount of $\mu$ distortions is roughly $\langle\mu(x)\rangle \simeq  2 \times 10^{-8}$ \cite{Chluba:2012gq, Cabass:2016giw, Chluba:2016bvg}. We find that $\mu$ distortions are tracers of primordial perturbations in the range $50 \textrm{ Mpc}^{-1} \lesssim k_{\mu} \lesssim 1\times10^4  \textrm{ Mpc}^{-1}$ \cite{Chluba:2012gq, Cabass:2016giw, Chluba:2016bvg}. 

\subsection{$\mu$ cross-correlations}

Anisotropies of the CMB probe the primordial curvature perturbations of inflation. CMB anisotropies are typically decomposed into spherical harmonics. For example, the real-space temperature anisotropy field $\Theta(\hat{n})=\delta T(\hat{n}) / T$, can be decomposed into spherical harmonics $\Theta(\hat{n})=\sum_{\ell m} a_{\ell m}^{T} Y_{\ell m}(\hat{n})$.
The spherical harmonic coefficients of the various CMB anisotropy fields are related to (Fourier-space) primordial curvature perturbations through
\begin{equation}
a^X_{\ell m} = 4\pi\, i^{-\ell}\int_{\vec{k}}e^{i\vec{k}\cdot\vec{x}}\zeta(\vec{k})\Delta^X_\ell(k)Y^\ast_{\ell m}(\vec{k})\,,
\end{equation}
where $X$ denotes the type of CMB anisotropy field (we are considering only $T$ and $E$ here), and $\Delta_{\ell}^X(k)$ is the transfer function connecting primordial perturbations to CMB anisotropies. 
The coefficients can then be correlated with each other:
\begin{equation}
    \left\langle\left(a_{\ell m}^{X}\right)^{*} a_{\ell^{\prime} m^{\prime}}^{Y}\right\rangle=\delta_{\ell \ell^{\prime}} \delta_{m m^{\prime}} C_{\ell}^{X Y},
\end{equation}
where $C_{\ell}^{X Y}$ is the angular cross-power spectrum of CMB anisotropy fields $X$ and $Y$. 

Observations of $\mu$-distortion anistropies can also be decomposed into spherical harmonics: 
\begin{equation}
\label{eq:forecast_app-A}
a^\mu_{\ell m} = 4\pi\, i^{-\ell}\int_{\vec{k}}e^{i\vec{k}\cdot\vec{x}}\mu(\vec{k})\Delta^\mu_\ell(k)Y^\ast_{\ell m}(\vec{k})\,,
\end{equation}
where $\Delta^\mu_\ell(k)$ is the transfer function of anisotropic $\mu$ distortions \cite{Cabass:2018jgj}:

\begin{eqnarray}
\Delta_{\ell}^{\mu}(k) & = & e^{-k^{2}/\left(q_{\mu,\mathrm{D}}^{2}\left(z_{\mathrm{rec}}\right)\right)} j_{\ell}(k \Delta \eta)  \\
\nonumber \Delta \eta &\equiv& \eta_{0}-\eta_{\mathrm{rec}} \\
\nonumber q_{\mu, \mathrm{D}}\left(z_{\mathrm{rec}}\right) & \simeq& 0.084 \, \mathrm{Mpc}^{-1},
\end{eqnarray}
and $\eta_0$ and $\eta_\mathrm{rec}$ are the conformal times corresponding to $z=0$ and the redshift of recombination $z=z_\mathrm{rec}$, respectively.
 The angular correlation of $\mu$ distortions with CMB anisotropies is then given by
\begin{eqnarray}
\label{eq:amut}
&&\left\langle a_{\ell m}^{\mu}\left(a_{\ell^{\prime} m^{\prime}}^{X}\right)^{*}\right\rangle = \delta_{\ell \ell^{\prime}} \delta_{m m^{\prime}} C_{\ell}^{\mu X} = \\
\nonumber &&(4 \pi)^{2} i^{-\ell+\ell^{\prime}} \int_{\vec{k}_\mathrm{S}} \int_{\vec{k}_\mathrm{L}} e^{i(\vec{k}_\mathrm{S}-\vec{k}_\mathrm{L}) \cdot \vec{x}} \times \\
\nonumber &&\left\langle\mu\left( \vec{k}_\mathrm{S}\right) \zeta(-\vec{k}_\mathrm{L})\right\rangle \Delta_{\ell}^{\mu}(\vec{k}_\mathrm{S}) \Delta_{\ell^{\prime}}^{X}(\vec{k}_\mathrm{L}) Y_{\ell m}^{*}(\hat{{k}_\mathrm{S}}) Y_{\ell^{\prime} m^{\prime}}(\hat{{k}_\mathrm{L}}),
\end{eqnarray}
where we have explicitly indicated here that, because of the very different transfer functions, this correlation probes the connection of the curvature power on very small scales through $\mu(\vec{k}_\mathrm{S}$) with the large-scale curvature $\zeta(\vec{k}_\mathrm{L})$. In other words, this measurement is sensitive to the correlation of a large-scale mode with two extremely small-scale modes, i.e., the bispectrum in the ultra-squeezed limit. 
The non-Gaussianities being probed are at scales of  $50 \textrm{ Mpc}^{-1} \lesssim k_\mu \lesssim 1\times10^4  \textrm{ Mpc}^{-1}$, much smaller than scales probed by \planck measurements of the primary CMB \cite{Planck:2019kim}. 

\begin{figure}[t!]
\centering
\includegraphics[width=1\columnwidth]{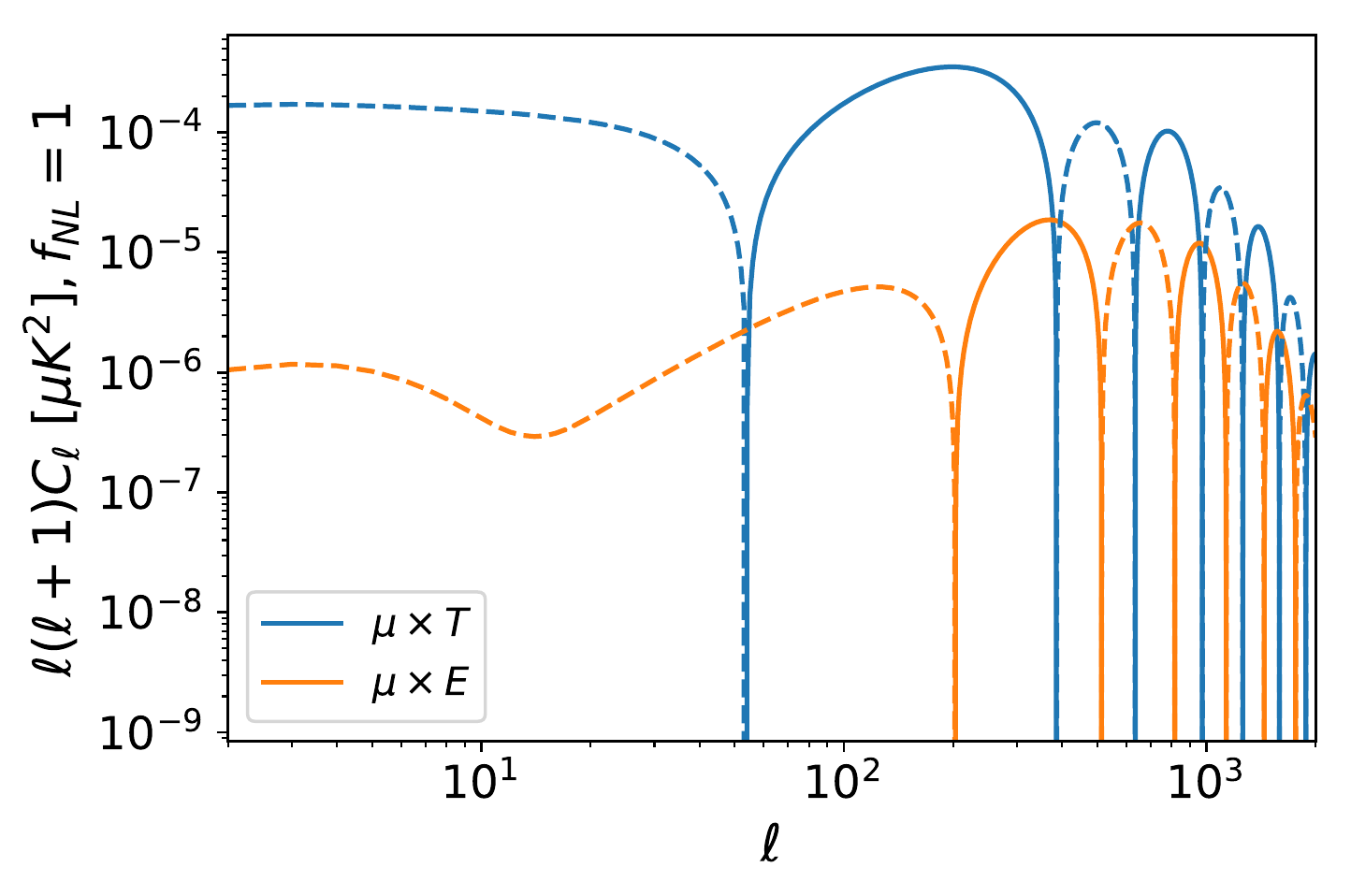}
\caption{ Angular cross-power spectrum of $  \mu \times T$ and $\mu \times E$ for $\langle\mu\rangle = 2 \times 10^{-8}$ and $\fnl = 1$. Solid lines correspond to positive values while dashed lines correspond to negative values.
\label{fig:mux}}
\end{figure}

The ensemble average of $\left\langle\mu\left( \vec{k}_\mathrm{S}\right) \zeta(-\vec{k}_\mathrm{L})\right\rangle$ is 
\begin{equation}
\label{eq:forecast_app-D}
\left\langle\mu\left( \vec{k}_\mathrm{S}\right) \zeta(-\vec{k}_\mathrm{L})\right\rangle=\left\langle\mu\right\rangle P_{\zeta}(k_\mathrm{L}) \frac{12}{5}\fnl.
\end{equation}
We use this to re-express $C_{\ell}^{\mu X}$ as 
\begin{equation}
C_{\ell}^{\mu X}  = \frac{24 \langle\mu\rangle}{5 \pi } \fnl  \int_{0}^{+\infty} \mathrm{d} k  P_{\zeta}(k) k^2 \Delta_{\ell}^{\mu}(k) \Delta_{\ell}^{X}(k) .
\label{eqn:clmux}
\end{equation}
The angular power spectra $C_{\ell}^{\mu T}$ and $C_{\ell}^{\mu E}$ for $\langle \mu \rangle = 2 \times 10^{-8}$ and $\fnl = 1$ are plotted in Figure~\ref{fig:mux}.

\subsubsection{$\mu \times T$ in single-field inflation}
\label{sec:sfi}
As discussed in Section~\ref{sec:powspec}, even in single-field inflation, squeezed-limit non-Gaussianity can be generated at the level of $\fnl \sim 1 - n_s$, where $n_s$ is the spectral index of the curvature power spectrum \cite{Maldacena:2002vr}. The mechanism that produces this squeezed-limit non-Gaussianity, however, produces vanishingly small amounts (far smaller than the level of $\fnl = 1 - n_s$) of $\mu \times T$ correlations. For this reason, $\mu \times T$ measurements have the potential to rule out single-field inflation more stringently than other probes. This was recognized by the authors of Ref.~\cite{Cabass:2018jgj}; we briefly summarize their argument here.

In single-field inflation, squeezed-limit non-Gaussianity is produced by long-wavelength modes that are outside of the Hubble radius modulating the global small scale power spectrum by acting as a modulation of the global scale, or a coordinate transformation, as seen in Eq.~(\ref{eq:modulate}). For $\mu$ distortions, this results in a shift in the observed value of $\mu$ in direction $\hat{n}$ of the form:
\begin{equation}
\mu\left(\hat{n}\right)=\mu\left(z_f, \vec{x}_{\mathrm{rec}}\right)+\zeta_L\left(z_f, \vec{x}_{\mathrm{rec}} \right)\hat{n}  \cdot \nabla_{\hat{{n}}} \mu\left(z_f, \vec{x}_{\mathrm{rec}}\right),
\end{equation}
where $\vec{x}_\mathrm{rec} = \hat{n} (\eta_0 - \eta_\mathrm{rec})$. 
However, the value of $\mu$ distortion at a given position is a local phenomenon and only traces the amount of power dissipated, which in turn depends on the value of the local small-scale power spectrum, which is invariant in the case of no other form of modulation. Thus, $\mu(\vec{x}) = \langle \mu \rangle$, and, relating the long wavelength mode $\zeta_L$ to the large-scale temperature anisotropy via $\zeta_L = - \frac{9}{2}\Theta_L$, in the Sachs-Wolfe approximation \cite{Cabass:2018jgj}, we can write
\begin{eqnarray}
\mu\left(\hat{n}\right) & \sim & \langle \mu \rangle+ \Theta_L   \hat{n} \cdot \nabla_{\hat{{n}}} \langle \mu \rangle, \\
\nonumber & = & \langle \mu \rangle + 0, 
\end{eqnarray}
This makes it clear that $\mu$ distortions from diffusion damping cannot correlate with a long wavelength temperature mode via this mechanism: 
\begin{equation}
    \langle\mu T\rangle \sim\left\langle\Theta_\mathrm{L}\right\rangle\left\langle\mu\right\rangle+\left\langle\Theta_\mathrm{L}^{2}\right\rangle \nabla_{\hat{n}}\left\langle\mu\right\rangle=\left\langle\Theta_\mathrm{L}\right\rangle\left\langle\mu\right\rangle=0.
\end{equation}
This means that $\mu \times T$ is vanishingly small from non-Gaussianity in single-field inflation, and the only terms that survive are ones with $(k_\mathrm{L}/k_\mathrm{S})^2$ suppression \cite{Cabass:2018jgj}. 
Therefore, any measurable signal of $\mu \times T$ would rule out single-field models. This similarly applies to $\mu \times E$.

\section{Methods}
\label{sec:methods}

Our aim is to forecast constraints on \fnl\ using joint analysis of  $\mu \times T$ and $\mu \times E$ correlations from CMB-S4. 
In this section, we will describe the key techniques we use to forecast sensitivity. First we describe the Fisher-matrix formalism used to convert $\mu$, $T$, and $E$ power spectra to constraints on \fnl. We then describe the component-separation technique that allows us to predict $\mu$ and $T$ or $E$ power spectra given the total signal + noise + foreground covariance matrix for a particular experimental configuration. Finally, we describe how we model the noise and foreground contributions to the band-band covariance matrix.

\subsection{Fisher Matrix}
\label{sec:fisher}

If we would like to know how accurately we can measure a given parameter $p_{i}$ in a data set, we can assume the likelihood $\mathcal{L}$ of measuring the parameter follows a Gaussian distribution:
\begin{eqnarray}
    \mathcal{L} \propto \exp \left[-\frac{1}{2}\left(p_{i}-\hat{p}_{i}\right) F_{ij}\left(p_{j}-\hat{p}_{j}\right)\right],
\end{eqnarray}
where $p_i$ is a fiducial value of the parameter and $\hat{p}_{i}$ is the measured value. $F_{ij}$ is the Fisher matrix, which captures the covariance of measured parameters. We discuss the validity of the assumption of Gaussian likelihood in Section~\ref{sec:compsep}.

We make the approximation that the amplitude of the $\mu \times T$ and $\mu \times E$ spectra are controlled by a single free parameter \fnl\ and write
\begin{eqnarray}
    C_{\ell}^{\mu X} = \left.\fnl C_{\ell}^{\mu X}\right|_{\fnl=1}.
\end{eqnarray}
The Fisher ``matrix'' in this case is a scalar:
\begin{eqnarray}
    F_{i j} & = &-\frac{\partial^{2} \ln \mathcal{L}}{\partial p_{i} \partial p_{j}} \\
    \nonumber &=& -\frac{\partial^{2} \ln \mathcal{L}}{\partial \fnl^2 } \delta_{ij} \equiv F, 
    \label{eqn:fisher}
\end{eqnarray}
and the 1$\sigma$ uncertainty on \fnl\ is 
\begin{equation}
    \sigma(\fnl) = \frac{1}{\sqrt{F}}.
\end{equation}

The expected noise at a given multipole $\ell$ is given by
\begin{eqnarray}
    \label{eqn:noise}
    \sigma_{\ell}^{2} &=& \left\langle\left(C_{\ell}^{\mu X}\right)^{2}\right\rangle-\left\langle C_{\ell}^{\mu X}\right\rangle^{2} \\
    \nonumber &=& \frac{C_{\ell}^{\mu \mu}  C_{\ell}^{XX} + (C_{\ell}^{\mu X})^2}{(2 \ell+1) \fsky},
\end{eqnarray}
where $\fsky$ is the fraction of the full sky observed in the survey. Our fiducial model for forecasting is $C_{\ell}^{\mu X} = 0$, and we neglect the contribution of the noise part of $C_{\ell}^{\mu X}$ to the variance, 
as it will always be much smaller than the product of $C_{\ell}^{\mu \mu}$ and $C_{\ell}^{XX}.$
The likelihood is thus
\begin{equation}
    -2 \ln \mathcal{L} = \sum_{\ell=\lmin}^{\ell_{\max }}(2 \ell+1) \fsky  \frac{\left(\left.\fnl C_{\ell}^{\mu X}\right|_{\fnl=1}\right)^{2}}{C_{\ell}^{\mu \mu} C_{\ell}^{XX}}.
\end{equation}

For the ${C_{\ell}^{\mu X}}$ cross-spectrum (the numerator or signal part of the Fisher matrix calculation), 
we use the formulation in Eq.~(\ref{eqn:clmux}).
For our fiducial forecasts, we adopt $\langle\mu\rangle = 2  \times 10^{-8}$. 
For the denominator or noise part of the Fisher calculation, we note that the auto-power-spectra ${C_{\ell}^{XX}}$ and ${C_{\ell}^{\mu \mu}}$
can be separated into signal and noise terms
\begin{eqnarray}
    {C_{\ell}} = {C_{\ell}^{S}} + {C_{\ell}^{N}},
\end{eqnarray}
where $S$ and $N$ denote signal and noise, respectively.
At CMB-S4 noise levels, measurements of both  ${C_{\ell}^{TT}}$ and ${C_{\ell}^{EE}}$ are signal-dominated for $\ell \leq 2000$ (at which point our constraints on \fnl\ are well saturated, see Figure~\ref{fig:fnl}). Therefore, we neglect the effects of noise and foregrounds on our temperature and $E$-mode anisotropy maps, and can make the approximation ${C_{\ell}^{XX}} \approx {C_{\ell}^{XX,S}}$. 
On the other hand, measurements of ${C_{\ell}^{\mu \mu}}$ will be noise-dominated for the foreseeable future, such that ${C_{\ell}^{\mu \mu}} \approx {C_{\ell}^{\mu \mu,N}}$. 
Calculations for ${C_{\ell}^{XX,S}}$ are taken from CAMB\footnote{http://camb.info} \cite{Lewis:1999bs}, while ${C_{\ell}^{\mu \mu, N}}$ is dependent on instrument and observation parameters, including instrumental and atmospheric noise levels. We describe how we obtain ${C_{\ell}^{\mu \mu, N}}$ from noise and foreground models in the following sections.

The constraining power of $C^{\mu T}_\ell$ and $C^{\mu E}_\ell$ for measuring $\fnl$ are comparable to each other. Rather than having independent constraints on $\fnl$ from $C^{\mu T}_\ell$ or $C^{\mu E}_\ell$, we can use the fact that they both probe the same underlying correlator $\langle\mu \zeta\rangle$ and perform a joint analysis. The authors of Ref.~\cite{Ravenni:2017lgw} and Ref.~\cite{Remazeilles:2021adt} demonstrated that the differing behavior of $C^{\mu T}_\ell$ and $C^{\mu E}_\ell$ with $\ell$ provides a better constraint on $\fnl$ than an independent analysis. We follow their method and modify the likelihood function to include the correlations between $T$ and $E$ in both the signal and the covariance. The final likelihood is then
\begin{eqnarray}
&& -2 \ln \mathcal{L} \approx \sum_{\ell=\lmin}^{\ell_{\max }}\frac{(2 \ell+1) \fsky}{C_{\ell}^{\mu \mu, N}\left[C_{\ell}^{T T} C_{\ell}^{E E}-\left(C_{\ell}^{T E}\right)^{2}\right]} \times \\
\nonumber && \bigg [ C_{\ell}^{T T}\left(\fnl C_{\ell}^{\mu E}|_{\fnl=1} \right)^{2}+C_{\ell}^{E E}\left(\fnl C_{\ell}^{\mu T}|_{\fnl=1}\right)^{2}- \\
\nonumber && 2 C_{\ell}^{T E} \fnl^2 C_{\ell}^{\mu T}|_{\fnl=1} C_{\ell}^{\mu E}|_{\fnl=1} \bigg ].
\label{eqn:fulllike}
\end{eqnarray}

\subsection{Component separation}
\label{sec:compsep}

For the purposes of our forecasting, observed maps of the total intensity of the sky in direction $\hat{n}$ and frequency $\nu_i$, $I_{i}(\hat{n})$, can be described as a linear combination of the CMB $\mu$ and $T$ signals and a noise contribution:
\begin{eqnarray}
\label{eqn:xtheta}
    I_{i}(\hat{n})=a_{\mu, i} s_{\mu}(\hat{n})+a_{\mathrm{T}, i} s_{\mathrm{CMB}}(\hat{n})+n_{i}(\hat{n}),
\end{eqnarray}
where $a_{\mu, i}$ and $a_{\mathrm{T}, i}$ are the $\mu$ and temperature spectral energy distributions (SEDs) at different frequency bands $i$, and $s_{\mu}(\hat{n})$ and $s_{\mathrm{CMB}}(\hat{n})$ are the true underlying $\mu$ and temperature anisotropy maps. 
We treat all astrophysical signals that are not
CMB $\mu$ or $T$ as noise and include them in $n$.
We can choose to work in spherical harmonic space, 
defining $I_{\ell m}$ such that $x(\hat{n}) = \sum_\ell \sum_m I_{\ell m} Y_{\ell m}(\hat{n})$. We can then rewrite Eq.~(\ref{eqn:xtheta}) as
\begin{eqnarray}
\label{eqn:xlm}
    I_{\ell m, i}=a_{\mu, i} s_{\ell m, \mu}+a_{\mathrm{T}, i} s_{\ell_m, \mathrm{CMB}}+n_{\ell m, i}.
\end{eqnarray}

Traditionally, data from CMB experiments are calibrated
such that maps in all frequency bands have the same 
response to primary CMB temperature anisotropy---i.e., the
maps are in units of CMB fluctuation temperature $\Delta T$ or fractional CMB fluctuation $\Delta T / T$.
In the latter case, 
the CMB temperature SED $a_{\mathrm{T}, i}$
is given by  
\begin{eqnarray}
    a_{\mathrm{T},i} = T_{\mathrm{CMB}}
\label{eq:exT}
\end{eqnarray}
for all bands.
We follow Ref.~\cite{Remazeilles:2018kqd} and approximate the $\mu$-distortion SED at frequencies of 20\,GHz and above as 
\begin{equation}
a_{\mu,i} = T_{\mathrm{CMB}}\left(\frac{1}{2.19}-\frac{1}{x_i}\right),
\label{eq:exmu}
\end{equation}
where
\begin{equation}
x \equiv \frac{h \nu}{k_\text{B} T_\text{CMB}}.
\end{equation}

We can obtain a temperature-free $\mu$ map or its spherical harmonic transform through component separation using a constrained internal linear combination method (CILC) \cite{Remazeilles:2010hq}. This method takes advantage of the known SEDs of temperature and $\mu$-distortion anisotropies and calculates weights $\boldsymbol{w}$ 
that, when applied to observed frequency maps, result in a $T$-free $\mu$ map and a $\mu$-free $T$ map. For example, if we assign the $\mu$ weights to the $i=0$ component of $\boldsymbol{w}_{ij}$, then the $T$-free $\mu$ map is given by
\begin{eqnarray}
    \hat{\mu}_{\ell m}^{T\text{-free}} & = & \sum_i \boldsymbol{w}_{0i} I_{\ell m, i}\\
    \nonumber & = & 1*s_{\mu}+0*s_{\mathrm{CMB}}+ \sum_i \boldsymbol{w}_{0i} n_{\ell m, i}
\end{eqnarray}
It is shown in Ref.~\cite{Remazeilles:2018kqd} that the weights that enforce unit response to $\mu$ distortions and zero response to temperature anisotropy, and minimize total variance, are given by 
\begin{equation}
\boldsymbol{w}^{T}=\boldsymbol{e}^{\mathrm{t}}\left(\mathrm{A}^{T} \mathbf{C}^{-1} \mathrm{~A}\right)^{-1} \mathrm{~A}^{T} \mathbf{C}^{-1} 
\label{eqn:lc}
\end{equation}
where 
\begin{eqnarray}
\mathrm{A} &=& \left(\begin{array}{lllll}
\boldsymbol{a}_{\mu} & \boldsymbol{a}_{\mathrm{CMB}} 
\end{array}\right)\\
\boldsymbol{e}^{T} &=& \left(\begin{array}{lllll}
1 & 0 
\end{array}\right),
\end{eqnarray}
$\mathbf{C} = \mathbf{C}^{ij}_{\ell}$ is the frequency-frequency $\ell$-space covariance matrix, and $T$ denotes transpose. 
The above weights can be generalized to null additional components $\boldsymbol{b}_{m}$
by generalizing $\mathrm{A} \rightarrow \left(\begin{array}{lllll}
\boldsymbol{a}_{\mu} & \boldsymbol{a}_{\mathrm{CMB}} & \boldsymbol{b}_{1} & \ldots & \boldsymbol{b}_{m}
\end{array}\right)$ and 
$\boldsymbol{e}^{T} \rightarrow \left(\begin{array}{lllll}
1 & 0 & 0 & \ldots & 0
\end{array}\right) $.
We note that the weights depend on multipole number $\ell$, i.e., $\boldsymbol{w} \rightarrow \boldsymbol{w}_\ell$. For simplicity of notation, we leave the $\ell$ dependence implicit and continue to use $\boldsymbol{w}$. 

The reduction of the full 
$\mathbf{C}^{ij}_{\ell m \ell^\prime m^\prime}=\left\langle I^{i}_{\ell m} I^{j}_{\ell^\prime m^\prime} \right\rangle$ to $\mathbf{C}^{ij}_{\ell}$ rests on the assumption that 
all sources of variance in the maps are isotropic, stationary, and Gaussian.
With detector noise only and assuming no correlations between detector noise at different frequencies, 
the covariance matrix is diagonal ($\mathbf{C}^{\mathrm{ij}}_\ell = \mathbf{C}^{\mathrm{ii}}_\ell \delta_{ij}$). Including foregrounds and atmosphere 
introduces correlated fluctuations between frequency bands and requires the full frequency-frequency matrix.

The assumption of Gaussianity and statistical isotropy is reasonably satisfied by detector noise, atmospheric emission, and extragalactic foregrounds, but not especially well by Galactic foregrounds. The CMB-S4 ``ultra-deep'' survey, which is the main survey we present forecasts for in this work (see Section~\ref{sec:surveys}) is located in an area of the sky with very low Galactic foreground emission. Furthermore, non-Gaussianity in the true foreground emission not accounted for in the covariance will not bias the component separation, it will only make it slightly suboptimal. We note that the Gaussianity of the likelihood (Eq.~\ref{eqn:fulllike}) depends on the power spectrum of the noise sources being Gaussian, not the pixel values or modes of the noise sources themselves, which is a lighter burden owing to the Central Limit Theorem.

We can then obtain ${C_{\ell}^{\mu \mu}}$ by applying the $\mu$-distortion weights to the frequency-frequency covariance matrix:
\begin{eqnarray}
    {C_{\ell}^{\mu \mu}} =  \sum_{ij} \boldsymbol{w}_{0i}\mathbf{C}^{\mathrm{ij}}_{_\ell} \boldsymbol{w}_{0j}.
\end{eqnarray}
We discuss the various contributions to the frequency-frequency covariance matrix $\mathbf{C}^{\mathrm{ij}}_\ell$ in the following sections.

\subsection{Instrument noise}
Assuming detector noise that is white (uncorrelated between time samples), uniform sky coverage, and a Gaussian instrument beam or point-spread function, the statistics of the map noise for frequency band $i$ in spherical harmonic or $\ell$ space are given by
\begin{eqnarray}
    n^i_\ell = N^i e^{\ell^2 \theta_i^2  / (16 \ln(2))},
\end{eqnarray}
where $N_i$ and $\theta_i$ is are the white noise level in the map and the angular resolution (beam FWHM) for frequency band $i$.
The contribution to the frequency-frequency covariance matrix from this source will be
\begin{eqnarray}
\mathbf{C}^{\mathrm{ij,N}}_\ell &=& \langle n^i n^j \rangle \\
\nonumber &=& N^i N^j 
e^{\ell^2 \theta_i \theta_j  / (8 \ln(2))}
\end{eqnarray}
and, for detector noise that is uncorrelated between bands,
\begin{equation}
    \mathbf{C}^{\mathrm{ij,N}}_\ell = (N^i)^2 
    e^{\ell^2 \theta_i^2  / (8 \ln(2))} \delta_{ij},
\end{equation}
where $\delta_{ij}$ is the Kronecker delta. 
For white detector noise only, the $\mu$-distortion power spectrum reduces to
\begin{eqnarray}
    {C_{\ell}^{\mu \mu}} = \sum_i \boldsymbol{w}_{0i}^2 (N^i)^2 
    e^{\ell^2 \theta_i^2  / (8 \ln(2))}.
\end{eqnarray}

\subsection{Atmosphere}

A major source of noise that must be considered in 
ground-based observations of the CMB is the emission
from blobs of poorly mixed water vapor in the Earth's
atmosphere (i.e., clouds).
For detailed discussions of this effect and 
measurements of the impact at various sites, see, e.g.,
Refs.~\cite{Lay:1999qi}, \cite{Bussmann:2004mw}, and 
\cite{Morris:2021jed}.
The spectrum of cloud sizes is such that the noise 
power from this source is much larger at 
large angular scales, and it is often modeled as 
a power law in $\ell$ (e.g., Ref. \cite{CMB-S4:2020lpa}). 
The total detector + atmosphere noise power in 
frequency band $i$ can then be parameterized with three 
numbers, namely the white noise level $N_i$, the multipole
value at which the detector and atmosphere noise
levels are equal $\ell_{\mathrm{knee},i}$, and the power-law index
of the atmosphere noise $\alpha^{\mathrm{atmo},i}$. 
\begin{equation}
     \mathbf{C}^{\mathrm{ij,N}}_\ell = \left [ 1+ \left ( \frac{\ell_{\mathrm{knee},i}}{\ell} \right )^{\alpha^{\mathrm{atmo},i}} \right ] (N^i)^2 e^{\ell^2 \theta_i^2  / (8 \ln(2))} \delta_{ij}.
\label{eqn:uatmo}
\end{equation}

Implicit in the formulation of Eq.~(\ref{eqn:uatmo})
is the assumption that the atmospheric noise is 
uncorrelated between bands. In fact, nearly the opposite
is the case, at least for detectors for which the 
beam patterns mostly overlap at the height of the
atmospheric emission. 
For example, internal CMB-S4 analysis of data from the SPT-3G receiver on the South Pole Telescope (SPT, \cite{Carlstrom:2009um,SPT-3G:2021vps}) found that for detectors in different frequency bands but co-located in a focal-plane pixel, the long-timescale fluctuations in the time-ordered data were over 99\% correlated.
To model this, we can introduce an atmospheric
correlation parameter $\eta$ and rewrite the noise + 
atmosphere contribution to the frequency-frequency
covariance matrix as
\begin{eqnarray}
    \mathbf{C}^{\mathrm{ij,N}}_\ell &=&
    \left [1 + \left ( \frac{\ell_{\mathrm{knee},i}}{\ell} \right )^{\alpha^{\mathrm{atmo},i}} \right ] \times \\
    \nonumber && (N^i)^2
    e^{\ell^2 \theta_i^2  / (8 \ln(2))}
    \delta_{ij} + \\
    \nonumber && + \left ( \frac{\ell_{\mathrm{knee},i}}{\ell} \right )^{\alpha^{\mathrm{atmo},i}/2}
    \left ( \frac{\ell_{\mathrm{knee},j}}{\ell} \right )^{\alpha^{\mathrm{atmo},j}/2} \times \\
    \nonumber && N^i N^j 
    e^{\ell^2 \theta_i \theta_j  / (8 \ln(2))} \pi_{ij},
\label{eqn:catmo}
\end{eqnarray}
where $\pi_{ij} = \eta(1 - \delta_{ij})$. In Sec.~\ref{sec:results}, we will show forecasts using values of using values of $\eta$ ranging
from 0 to 1. As the correlation $\eta$ increases, the component separation algorithm defined in Sec.~\ref{sec:compsep} is more effective in reducing the atmospheric contribution to the final ${C_{\ell}^{\mu \mu}}$ covariance.

\subsection{Foregrounds}

In our forecasting pipeline, we also consider the effects of foregrounds. Foregrounds contribute to the frequency-frequency covariance matrix as linear additions (because they are not correlated with the other sources of covariance) such that
\begin{eqnarray}
\mathbf{C}^{\mathrm{ij}}_{\ell} =  \mathbf{C}^{\mathrm{ij,N}}_{\ell} + \mathbf{C}^{\mathrm{ij,fore}}_{\ell}
\end{eqnarray}
For this work, we approximate all foregrounds
as 100\% correlated across frequency bands (though we approximate each foreground source as uncorrelated with the others), such that for a given foreground
type (call it ``type X'')
\begin{equation}
    \mathbf{C}^{\mathrm{ij,X}}_{\ell} = \sqrt{C_\ell^{ii,\mathrm{X}} C_\ell^{jj,\mathrm{X}}}.
\end{equation}

\begin{figure}[t!]
\centering
\includegraphics[width=1\columnwidth]{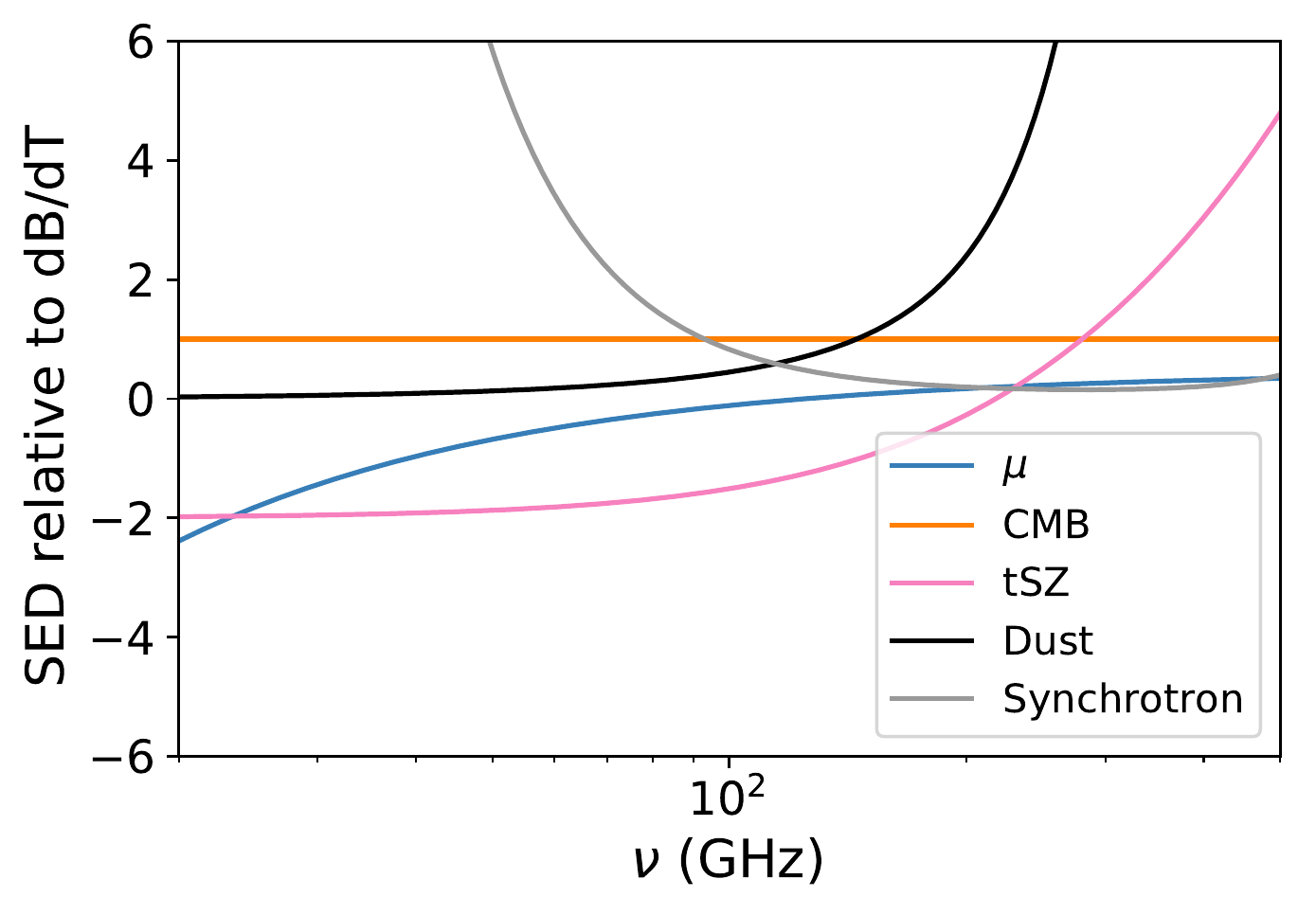}
\caption{Spectral shape of CMB anisotropy, $\mu$ distortions, and various Galactcic and extragalactic foregrounds, all divided by the CMB anisotropy SED and scaled by an arbitrary amplitude. For frequencies $\nu < 100$\,GHz, we see that Galactic synchrotron sharply rises, even relative to $\mu$ distortions. At frequencies $\nu > 100$\,GHz, Galactic dust and the thermal SZ effect start growing as the $\mu$-distortion spectrum starts leveling off. 
\label{fig:SED}}
\end{figure}

Foreground sources can be separated into Galactic and extragalactic sources, and we treat each of these in turn below.
Often in the literature, foreground amplitudes and
$\ell$-space behavior are quoted in 
$D_\ell = \frac{\ell(\ell+1)}{2 \pi} C_\ell$.
When we adopt such parameterizations, we keep the 
description in $D_\ell$ but convert to $C_\ell$
when actually implementing the model. 
We show the SEDs of the primary foreground sources, along with the $T$ and $\mu$ SEDs, in Figure~\ref{fig:SED}. 

\subsubsection{Galactic Sources}

The primary sources of Galactic contamination at CMB observing frequencies are thermal dust emission and synchrotron emission.
Interstellar dust heated by starlight emits as a quasi-thermal modified blackbody. We follow Ref.~\cite{CMB-S4:2020lpa} and parameterize the frequency behavior and $\ell$-space shape of thermal dust emission as
\begin{eqnarray}
\begin{aligned}
    D_{\ell, \nu_{1}, \nu_{2}}^\mathrm{d}=D_{80,\nu_0}^\mathrm{d}    \epsilon_{\nu_{1}, \nu_{2}}\left(\frac{\nu_{1} \nu_{2}}{\nu_{0}^{2}}\right)^{\alpha^\mathrm{d}} \times \\ 
    \frac{B_{\nu_1}\left(T_{\mathrm{d}}\right) B_{\nu_2}\left(T_{\mathrm{d}}\right)}{B^2_{\nu_0}\left(T_{\mathrm{d}}\right)}\left(\frac{\ell}{80}\right)^{-0.4},
\end{aligned}
\end{eqnarray}
where $B$ is Planck's law, 
$D^\mathrm{d}_{80,\nu_0}$ is the value of $D$ at $\ell=80$ and the reference frequency $\nu_0$, $T_\mathrm{d}$ is the dust temperature, and $\alpha^\mathrm{d}$ is the dust spectral index. Following Ref.~\cite{Dibert:2022gep},
we define $D_{80}^\mathrm{d}=3.253 \, \mu \mathrm{K}^{2}$, $\alpha^\mathrm{d} = 1.6$, $T_\mathrm{d} = 19.6$, and $\nu_0 = 145$\,GHz.
Finally, 
$\epsilon_{\nu_{1}, \nu_{2}}$ relates the brightness
of CMB fluctuations at $\nu_1$, $\nu_2$, and $\nu_0$:
\begin{eqnarray}
    \epsilon_{\nu_{1}, \nu_{2}} \equiv \frac{\left[\frac{d B(\nu_0,T_\mathrm{CMB})}{d T}\right]^2}
   {\frac{d B(\nu_1,T_\mathrm{CMB})}{d T} \frac{d B(\nu_2,T_\mathrm{CMB})}{d T}}.
\end{eqnarray}
Because we have defined our $\mu$ and $T$ SEDs assuming that 
the input maps are calibrated to CMB fluctuation 
amplitudes, we must also account for this in the 
foreground modeling.
The ratio $1/(dB/dT |_{\nu,T_\mathrm{CMB}})$ converts the source radiance of a given foreground at frequency $\nu$ to an equivalent CMB temperature anisotropy.

We use a similar parameterization for Galactic synchrotron, again following Ref.~\cite{CMB-S4:2020lpa}
\begin{eqnarray}
    D_{\ell, \nu_{1}, \nu_{2}}^\mathrm{sync}=D_{80,\nu_0}^\mathrm{sync} \epsilon_{\nu_{1}, \nu_{2}} \left(\frac{\nu_{1} \nu_{2}}{\nu_{0}^{2}}\right)^{\alpha^\mathrm{sync}}\left(\frac{\ell}{80}\right)^{-0.4},
\end{eqnarray}
and we again adopt values from Ref.~\cite{Dibert:2022gep}:  $D_{80}^\mathrm{sync}=0.005 \, \mu \mathrm{K}^{2}$, $\alpha^\mathrm{sync} = -1.10$, and $\nu_0 = 93$\,GHz. We note that both the synchrotron and dust amplitudes used here are estimated for the CMB-S4 ``ultra-deep'' $\fsky=0.03$ survey (see next section for details).

Because of the shape of the $\mu$-distortion SED, lower-frequency channels are particularly important for recovering the signal, and it is possible that other Galactic foregrounds such as free-free emission and ``anomalous  microwave  emission'' (AME) could be important contaminants. We investigate the behavior of these additional foregrounds in the CMB-S4 3\% sky region using {\tt PySM} \cite{Thorne:2016ifb}, which is based on the \textit{Planck} Sky Model \cite{Delabrouille:2012ye}. 
We find that the AME SED has a double-peaked shape, which we parameterize as
\begin{eqnarray}
&& f^2_\mathrm{AME}(\nu) = \\ 
\nonumber && \frac{e^{-\left[\ln(\nu) -\ln(\nu_{a1})\right]^2/2 \sigma_{a1}^2} + A e^{-\left[\ln(\nu) -\ln(\nu_{a2})\right]^2/2 \sigma_{a2}^2}}{1 + A e^{-\left[\ln(\nu_{a1}) -\ln(\nu_{a2})\right]^2/2 \sigma_{a1}^2 }}, 
\end{eqnarray}
with $\nu_{a1} = 10\,\mathrm{GHz}$, $\sigma_{a1} = 0.43\,\mathrm{GHz}$, $\nu_{a2} = 22\,\mathrm{GHz}$, $\sigma_{a2} = 0.35\,\mathrm{GHz}$, and $A = 0.0065$. We assume similar $\ell$-space behavior as the thermal dust and write
\begin{equation}
D_{\ell, \nu_{1}, \nu_{2}}^{\mathrm{AME}} = D_{80, \nu_0}^{\mathrm{AME}} f_\mathrm{AME}(\nu_1) f_\mathrm{AME}(\nu_2) \left(\frac{\ell}{80}\right)^{-0.4},
\end{equation}
with $D_{80, \nu_0}^{\mathrm{AME}} = 1.0 \times 10^4 \mu \mathrm{K}^2$ at $\nu_0 = 10$\,GHz.
The free-free emission in the CMB-S4 3\% sky region appears to be dominated by point-like sources (either residual contributions from extragalactic radio sources or emission from compact HII regions). Because CMB-S4 will have the sensitivity and resolution to find and mask such sources, we neglect the contribution of free-free emission in this analysis. 

\subsubsection{Extragalactic foregrounds}
In addition to emission from our Galaxy, we also need to consider foreground emission from extragalactic sources. We treat four independent types of extragalactic foregrounds: the thermal Sunyaev-Zel'dovich (tSZ) effect, the clustered cosmic infrared background (CIB), the unclustered (``shot-noise'' or ``Poisson'') component of the CIB, and synchrotron-emitting active galactic nuclei (or ``radio sources''), the clustering of which is assumed to be negligible. We neglect the contribution from the kinetic Sunyaev-Zel'dovich (kSZ) effect, which has the same frequency spectrum as the CMB and will be nulled by the CILC.

The tSZ effect, a type of $y$ distortion, arises from CMB photons scattering off electrons in the intra-cluster
medium of galaxy clusters and other collapsed structures. This shifts the re-emitted photons to higher frequencies compared to the CMB spectrum. 
We parameterize the tSZ as
\begin{eqnarray}
    D_{\ell, \nu_{1}, \nu_{2}}^\mathrm{tSZ}=D^\mathrm{tSZ}_{3000,\nu_0} T(\ell) \frac{f(\nu_1)f(\nu_2)}{f^2(\nu_0)},
\end{eqnarray}
where
\begin{eqnarray}
    f(\nu)=x \frac{e^{x}+1}{e^{x}-1}-4,
\end{eqnarray}
\begin{eqnarray}
    x = \frac{h \nu}{k_\mathrm{B} T},
\end{eqnarray}
and $T(\ell)$ is the tSZ template used in Ref.~\cite{Story:2012wx}. Following Ref.~\cite{George:2014oba}, we adopt $D^\mathrm{tSZ}_{3000,\nu_0} = 3.4\,\mu$K$^2$ at $\nu_0 = 150$\,GHz.

Again following Ref.~\cite{George:2014oba}, we parameterize the clustered CIB as
\begin{eqnarray}
    D_{\ell, \nu_{1}, \nu_{2}}^\mathrm{c}=D_{3000}^\mathrm{c}(\nu_0) \epsilon_{\nu_{1}, \nu_{2}} \left(\frac{\nu_{1} \nu_{2}}{\nu_{0}^{2}}\right)^{\alpha^\mathrm{p}}\left(\frac{\ell}{3000}\right)^{0.8},
\end{eqnarray}
where
$D_{3000}^\mathrm{c}(\nu_0)=3.46 \mu \mathrm{K}^{2}$, $\alpha^\mathrm{p}$ = 4.27, and $\nu_0 = 150$\,GHz. 
And we parameterize the Poisson component of the CIB as
\begin{eqnarray}
    D_{\ell, \nu_{1}, \nu_{2}}^\mathrm{p}=D_{3000}^\mathrm{p}(\nu_0) \epsilon_{\nu_{1}, \nu_{2}} \left(\frac{\nu_{1} \nu_{2}}{\nu_{0}^{2}}\right)^{\alpha^\mathrm{p}}\left(\frac{\ell}{3000}\right)^{2},
\end{eqnarray}
with $D_{3000}^\mathrm{p}=9.16 \mu \mathrm{K}^{2}$, $\alpha^\mathrm{p}$ = 3.27, $\nu_0 = 150$\,GHz.
For this initial work, we do not consider spatial correlation between the tSZ and CIB. This does not cause any direct bias to our final result, as we explicitly null any signal with a tSZ spectrum in our final constraints (see Section~\ref{sec:results} for details). For more discussion of the effects of tSZ-CIB correlation on $\mu \times T$ measurements, see Ref.~\cite{Rotti:2022ifo}.

Finally, we parameterize radio source power as
\begin{eqnarray}
    D_{\ell, \nu_{1}, \nu_{2}}^\mathrm{r}=D_{3000}^\mathrm{r}(\nu_0) \epsilon_{\nu_{1}, \nu_{2}} \left(\frac{\nu_{1} \nu_{2}}{\nu_{0}^{2}}\right)^{\alpha^\mathrm{p}}\left(\frac{\ell}{3000}\right)^{2},
\end{eqnarray}
with $D_{3000}^\mathrm{r}(\nu_0)= 0.02 \, \mu \mathrm{K}^{2}$, $\alpha^\mathrm{p}$ = -0.7, $\nu_0 = 150$\,GHz.
This is significantly lower than amplitudes quoted in, e.g., Ref.~\cite{George:2014oba}.
This is because the radio Poisson power is dominated by the brightest individual sources in the map, and masking and removing the brightest radio sources will reduce the Poisson term. For an experiment similar to CMB-S4, which will achieve roughly 1 $\mu$K-arcmin map noise,  radio sources can be cleaned down to roughly 0.2\,mJy, which reduces their Poisson amplitude to $D_{3000}^\mathrm{r}(\nu_0)=0.02 \, \mu \mathrm{K}^{2}$.

\subsection{Calibration}
Historically, measurements of absolute brightness or temperature at microwave/millimeter-wave frequencies have been successfully carried out only by space- or balloon-borne telescopes, because of the stringent requirements on calibration accuracy and stability (e.g., Refs. \cite{Mather:1993ij, Kogut:2006kf}). Proposed measurements of the absolute $\mu$-distortion amplitude, such as with the PIXIE satellite \cite{Kogut:2010xfw}, are designed with similar constraints in mind. It was pointed out, however, in Ref.~\cite{Ganc:2012ae} that measuring the \textit{anisotropy} of $\mu$ distortions, and particularly the correlated anisotropy of $\mu$ and temperature, effectively converts the absolute calibration requirement to a relative calibration requirement, and the bias on the absolute measurement to an uncertainty on the measurement of anisotropy. 

A calibration error in CMB-S4-like data will result primarily in leakage of the much larger temperature signal (or foregrounds) into the component-separated $\mu$ map, resulting in a component of $T \times T$ in the $\mu \times T$ cross-spectrum. The $T \times T$ spectrum does not have the same shape as $\mu \times T$ (which crosses zero many times over the $\ell$ range we consider), so there will be on average no bias from this leakage, just excess variance. We have investigated the level to which the relative calibration must be known for this variance not to dominate the error on \fnl, but a simple scaling argument tells us that, because the relative calibration will come from enforcing equal response to the CMB temperature in every frequency band, the calibration precision in any one band will be equal to the S/N on the CMB temperature anisotropy in that band. Thus, the contribution to uncertainty on 
\fnl\ from calibration errors will be on the order of the contribution from noise divided by the square root of the number of bands.
For this reason, we ignore calibration uncertainty in our main results. 

A related concern is the knowledge of the instrumental bandpasses. Even in the limit of perfect relative calibration off the CMB, imperfect knowledge of the instrument bandpasses could lead to a different level of foreground contamination in the $\mu$ map than the component-separation algorithm predicts. As with CMB $T$ leakage, this will not generally have the same shape as the $\mu \times T$ spectrum and will thus not cause bias on average. Furthermore, as we are not explicitly projecting out foregrounds in the component separation, the extra leakage of foregrounds into the $\mu$ map from bandpass uncertainty is likely to be small. Nevertheless, we will update the forecasting machinery to include this effect in a future paper.

\section{Survey configuration}
\label{sec:surveys}

\begin{figure}[htbp]
\begin{subfigure}
    \centering
    \includegraphics[width=.95\columnwidth]{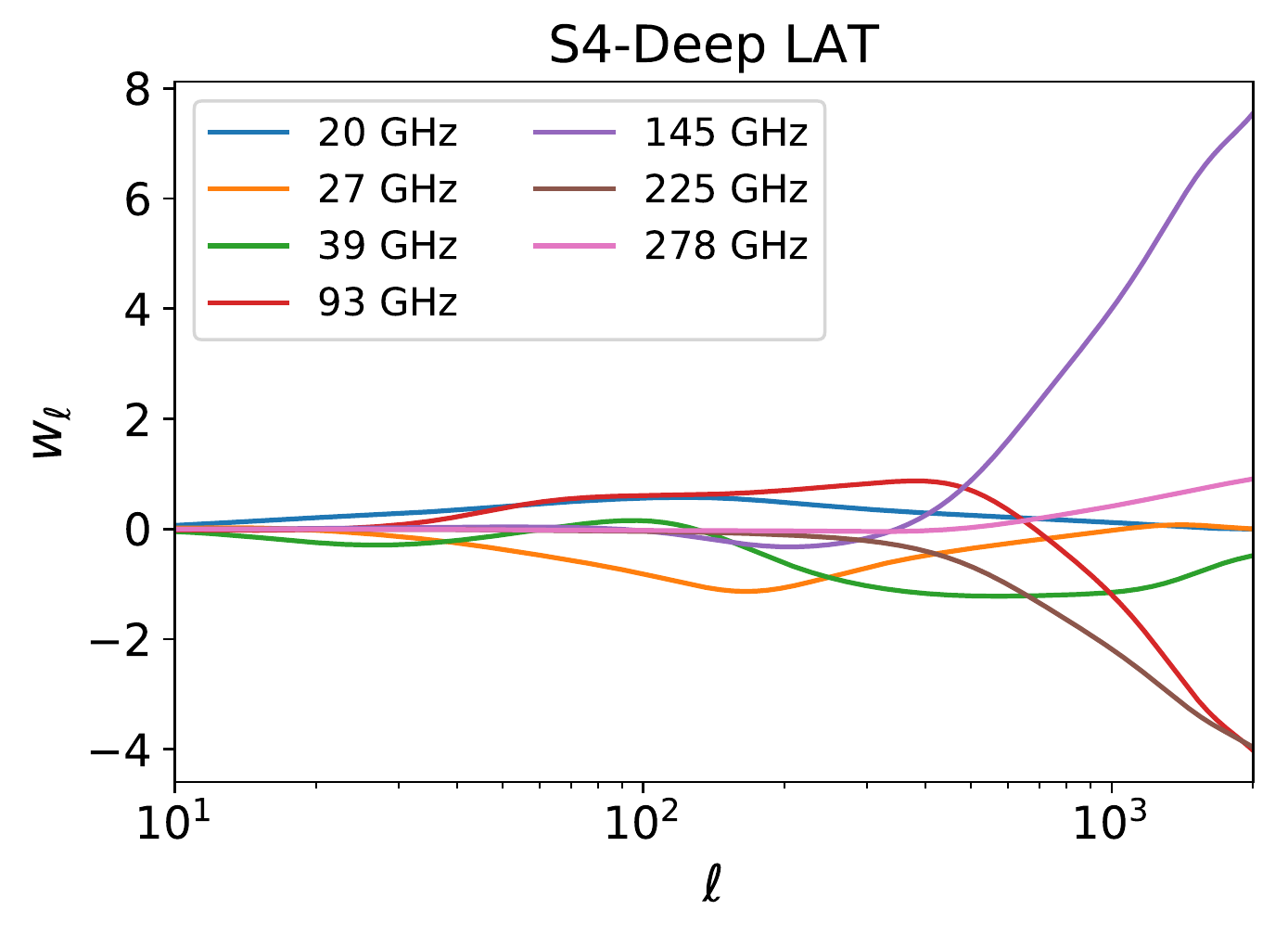}
\end{subfigure}
\begin{subfigure}
    \centering
    \includegraphics[width=.95\columnwidth]{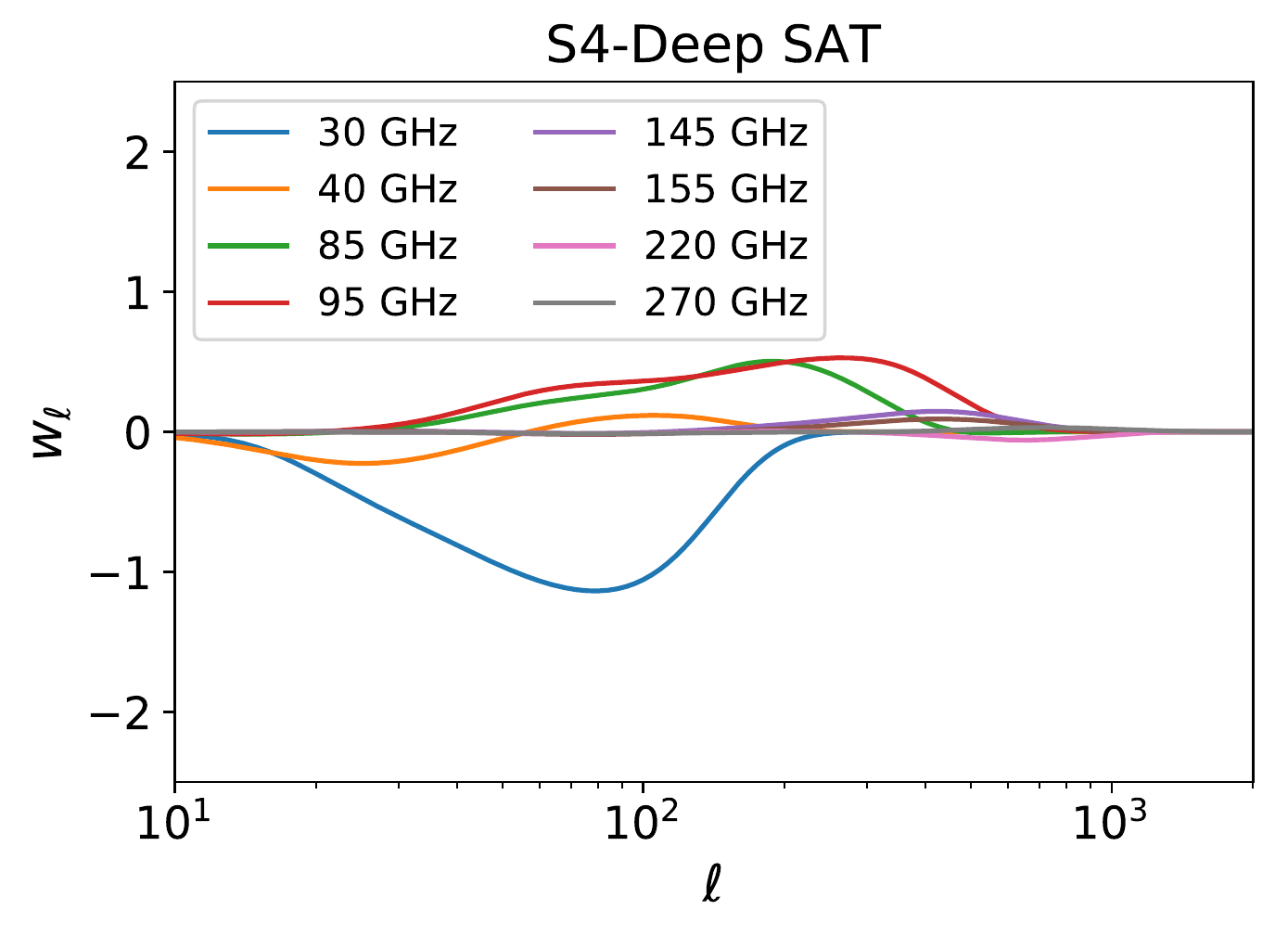}
\end{subfigure}
\begin{subfigure}
    \centering
    \includegraphics[width=.95\columnwidth]{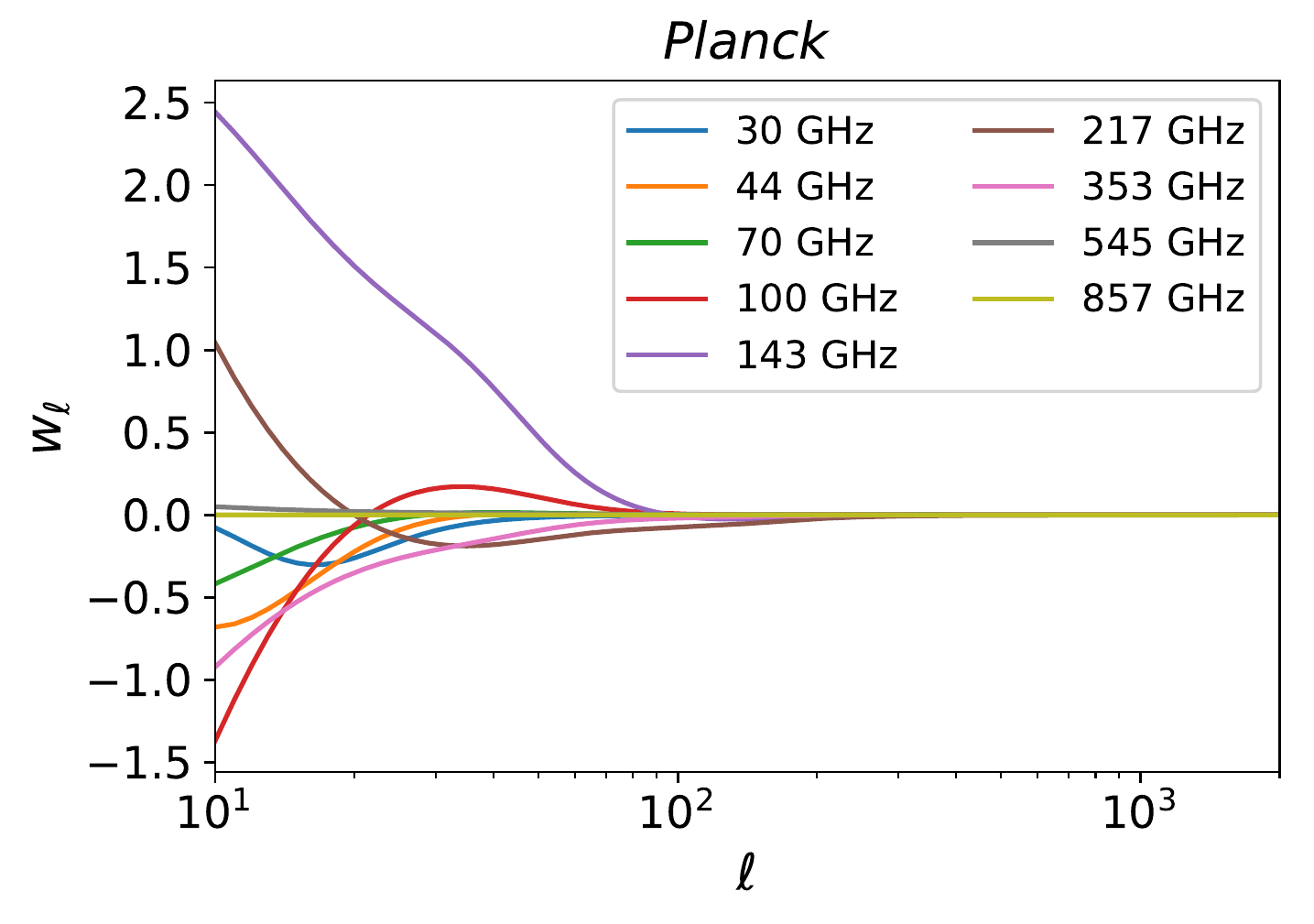}
\end{subfigure}
\caption{Weights used to construct the $T$-free $\mu$-distortion map (i.e., the $\mu$ component of  $\boldsymbol{w}$) for the CMB-S4 ultra-deep patch from different frequency channels, including foregrounds and assuming 99\% correlated atmosphere.
\label{fig:weights}}
\end{figure}

\begin{table*}[ht]
\def\arraystretch{1.5}
\setlength{\tabcolsep}{7pt}
\centering
\begin{tabular}{|l|ccccccc|}
\hline Frequency (GHz) & 20 & 27 & 39 & 93 & 145 & 225 & 278 \\
\hline Angular resolution (arcmin) & 11 & 8.4 & 5.8 & 2.5 & 1.6 & 1.1 & 1.0 \\
White noise level in temperature ($\mu \mathrm{K}$-arcmin) & 9.31 & 4.6 & 2.94 & 0.45 & 0.41 & 1.29 & 3.07 \\
\lknee\ for $TT$ & 400 & 400 & 400 & 1200 & 1900 & 2100 & 2100 \\
$\alpha$ for $TT$ & 4.2 & 4.2 & 4.2 & 4.2 & 4.1 & 4.1 & 3.9 \\
\hline
\end{tabular}
\caption{
Survey configuration for the large-aperture telescope (LAT) in the CMB-S4 ultra-deep survey.
}
\label{tab:s4lat}
\end{table*}

\begin{table*}[ht]
\def\arraystretch{1.5}
\setlength{\tabcolsep}{7pt}
\centering
\begin{tabular}{|l|cccccccc|}
\hline Frequency (GHz) & 30 & 40 & 85 & 95 & 145 & 155 & 220 & 270 \\
\hline Angular resolution (arcmin) & 72.8 & 72.8 & 25.5 & 22.7 & 25.5 & 22.7 & 13.0 & 13.0 \\
White noise level in polarization ($\mu \mathrm{K}$-arcmin) & 3.53 & 4.46 & 0.88 & 0.78 & 1.23 & 1.34 & 3.48 & 5.97 \\
White noise level in temperature ($\mu \mathrm{K}$-arcmin) & 2.50 & 3.15 & 0.62 & 0.55 & 0.87 & 0.95 & 2.46 & 4.22  \\
\lknee\ for $TT$ & 400 & 400 & 1200 & 1200 & 1900 & 1900 & 2100 & 2100 \\
$\alpha$ for $TT$ & 4.2 & 4.2 & 4.2 & 4.2 & 4.1 & 4.1 & 4.1 & 3.9 \\
\hline
\end{tabular}
\caption{
Survey configuration for the small-aperture telescopes (SATs) in the CMB-S4 ultra-deep survey.
}
\label{tab:s4sat}
\end{table*}

\begin{table*}[ht]
\def\arraystretch{1.5}
\setlength{\tabcolsep}{7pt}
\centering
\begin{tabular}{|l|ccccccccc|}
\hline Frequency (GHz) & 30 & 44 & 70 & 100 & 143 & 217 & 353 & 545 & 857 \\
\hline Angular resolution (arcmin) & 32.4 & 27.1 & 13.3 & 9.69 & 7.30 & 5.02 & 4.94 & 4.83 & 4.64 \\
White noise level for $TT$ ($\mu \mathrm{K}$-arcmin) & 150 & 162 & 210 & 77.4 & 33.0 & 46.8 & 154 & 815 & $2.98 \times 10^{4}$ \\
\hline
\end{tabular}
\caption{
Angular resolution and noise levels assumed for \planck (reproduced from Table 4 of  Ref.~\cite{Planck:2018nkj}). Note that we assume \planck noise is white down to the \lmin\ of our forecasting.
}
\label{tab:planck}
\end{table*}

In this work, we are forecasting $\sigma(\fnl)$ using the parameters of the upcoming CMB-S4 experiment \cite{Abazajian:2019eic}. There are two major surveys planned for CMB-S4, an ``ultra-deep'' survey of roughly 3\% of the Southern sky, conducted from the South Pole with both small-aperture telescopes (SATs) and a large-aperture telescope (LAT), and a ``deep and wide'' survey of roughly 60\% of the sky, conducted from Chile with LATs only.

Our fiducial forecasting will be for the $\fsky = 0.03$ survey.
The $\fsky=0.6$ survey involves a similar investment of total detector number and observing time, but spread across a sky patch that is 20 times larger.
In the limit of detector noise only, 
this introduces a factor \fsky\ into the covariance matrix, leading to
\begin{eqnarray}
    \mathbf{C}^{\mathrm{ij,N}}_{\ell} (\fsky) = \mathbf{C}^{\mathrm{ij,N}}_{\ell} (\fsky=0.03) \frac{\fsky}{0.03} .
\end{eqnarray}
Because incomplete sky surveys also observe fewer total
sky modes and hence have fewer samples of the various 
power spectra, 
the Fisher matrix calculations also include an \fsky\ term, as in Eq.~(\ref{eqn:noise}).

As discussed in the previous section, we can make the approximation that $C_\ell^{\mu \mu}$ is noise-dominated, while $C_\ell^{TT}$ is signal-dominated for $\ell \leq 2000$.
This means that $C_\ell^{\mu \mu}$ is the only term in the denominator of the $\mu \times T$ Fisher calculation that has an \fsky\ dependence:
\begin{eqnarray}
-2 \ln \mathcal{L} &=& \sum_{\ell=2}^{\ell_{\max }} 0.03 \ \left (2 \ell+1 \right) \frac{\fsky}{0.03} \ \times \\
\nonumber && \frac{\left(\left.\fnl C_{\ell}^{\mu T}\right|_{f_\mathrm{NL}}\right)^{2}}{(\fsky/0.03) \ C_{\ell}^{\mu \mu, \mathrm{N}}|_{f_\mathrm{sky}=0.03} \ C_{\ell}^{T T}}.
\end{eqnarray}
In this limit,  
\fsky\ cancels out, leaving our Fisher matrix independent of \fsky\ in the detector-noise-only case. We note that this will be true for any measurement that depends on correlating a very small signal (that is below the detection threshold for a given experiment) with a much larger one (that is measured at high S/N).\footnote{One caveat to this is that the survey needs to be big enough to resolve the largest scale that is important for measuring the signal.}
When we add atmosphere and foregrounds, the situation becomes more complicated, because the statistics of the atmospheric noise are different at the two CMB-S4 sites, and because it is more difficult to avoid bright parts of our Galaxy when more sky is observed.

The parameters of the high-resolution (LAT) part of the ultra-deep ($\fsky=0.03$) survey, including band centers, angular resolutions, and noise levels, are given in Table~\ref{tab:s4lat}. The corresponding values for the degree-scale (SAT) ultra-deep survey are shown in Table~\ref{tab:s4sat}.

The $\fsky = 0.03$ ultra-deep survey for CMB-S4 will be conducted from the South Pole.  In terms of atmospheric emission at CMB frequencies, the South Pole is the best large, developed site on Earth \cite{Kuo:2017ubm}. 
In Tables \ref{tab:s4lat} and \ref{tab:s4sat}, we show the expected values
of \lknee\ and $\alpha$ (see Eq.~\ref{eqn:uatmo}) for the CMB-S4 bands at the South Pole, derived from CMB-S4 internal 
analysis of SPT-3G data.
We note that the Deep Survey LAT value for \lknee\ for bands below 40\,GHz in official CMB-S4 documents is 1200, while we use 400 here. The value of 1200 is a conservative choice made assuming no improvement in atmospheric noise from the lowest SPT band of 95\,GHz. The \lknee\ values for the Wide Survey LAT low-frequency bands, which were derived by scaling the atmospheric noise power with levels of precipitable water vapor, are close to 400, and thus we adopt that value for the Deep Survey here. We discuss the impact of this choice in Section~\ref{sec:results}.

In addition to reducing raw noise levels, the authors of Ref.~\cite{Remazeilles:2018kqd} suggest expanding detector frequency coverage, in order to lower $C^{\mu \mu,N}$. To increase the upper frequency range of CMB-S4 measurements, we include \planck 2018 data in the forecast, which extends up to 857\,GHz. 
This should provide valuable complementary information that can help mitigate foregrounds at frequencies that CMB-S4 is unable to observe.
We implement \planck bands in our forecasting pipeline in the same way as CMB-S4 (as independent frequency channels, over the same 3\% of the sky as CMB-S4 Deep), using the parameters shown in Table~\ref{tab:planck}. In Figure~\ref{fig:weights}, we show the components of $\boldsymbol{w}$ for the $T$-free $\mu$ map contributed from each individual CMB-S4 or \planck\ band, assuming 99\% atmospheric correlation and the foregrounds described in the previous section.

\begin{figure}[t!]
\centering
\includegraphics[width=1\columnwidth]{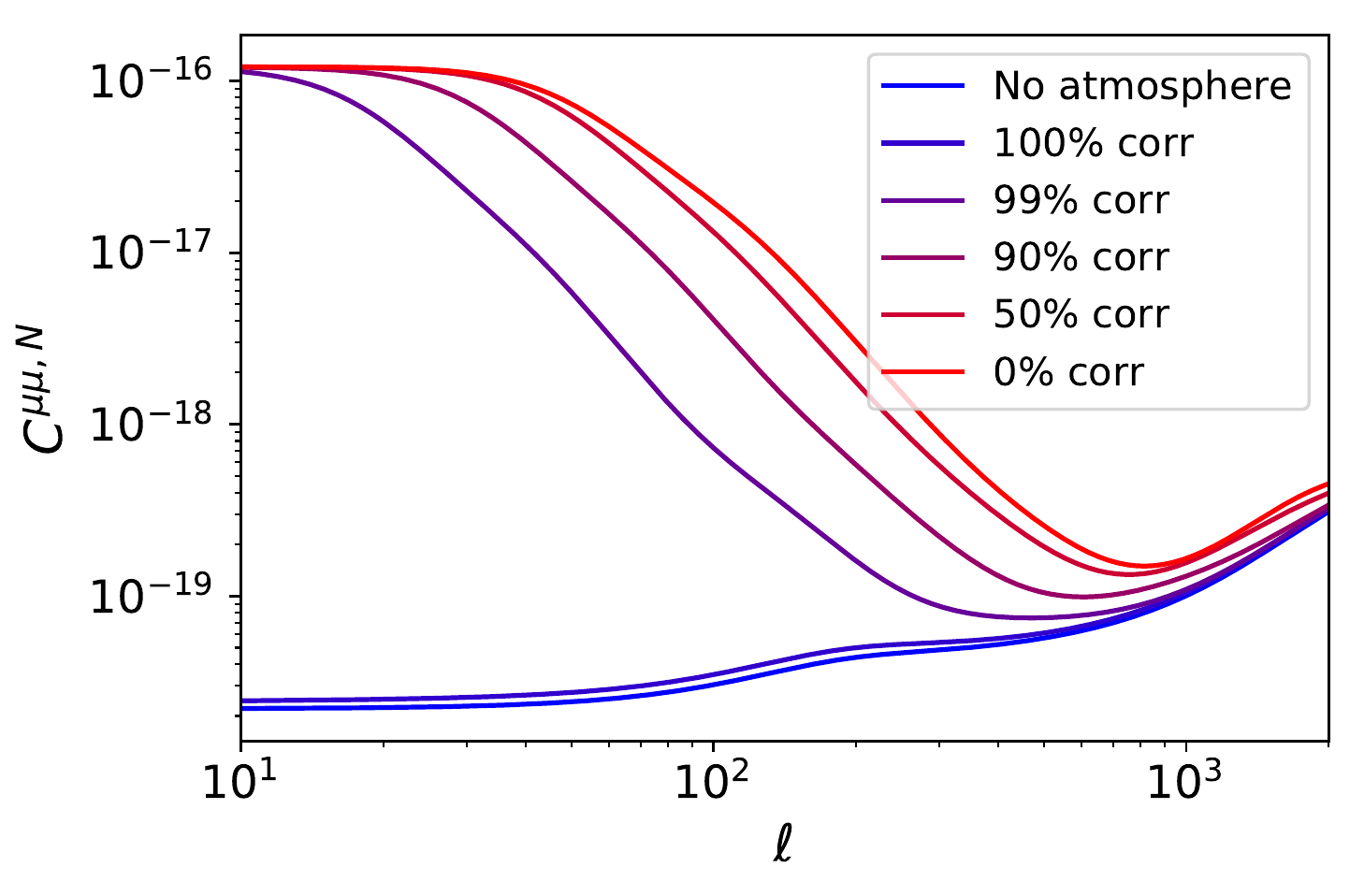}
\caption{$C^{\mu \mu, N}_\ell$ for the CMB-S4 ultra-deep survey and various assumptions for the degree of atmospheric correlation between bands. Full correlation among bands closely matches the nominal case of no atmosphere. However, even minimal decorrelation among detectors significantly worsens $C^{\mu \mu, N}_\ell$ at low multipoles.
\label{fig:mumu}}
\end{figure}

In Figure~\ref{fig:mumu}, we show the expected noise power in the $\mu$ map from CMB-S4 ($C^{\mu \mu, N}_\ell$) for various values of atmospheric correlation among frequency bands. In the scenario where frequency bands are fully correlated and the atmosphere can be fully subtracted out, the associated noise curve closely matches the scenario with no atmosphere. However, even minimal decorrelation among detectors significantly worsens $C^{\mu \mu, N}_\ell$ at low multipoles. Residual atmosphere acts as an effective \lmin\ when summing over multiple observation bands to constrain $\fnl$. 

We note that currently fielded experiments such as Advanced ACTPol \cite{Henderson:2015nzj} and SPT-3G, as well as the upcoming Simons Observatory (SO) experiment \cite{SimonsObservatory:2018koc} can be approximated in this forecasting framework as versions of either the CMB-S4 Deep or Wide surveys, but with higher noise and (in the case of currently fielded experiments) reduced frequency coverage. In particular, the SO LAT ``goal" survey is similar to the CMB-S4 Wide survey with 2-3 times higher noise and slightly less sky area, thus we would thus expect to forecast roughly 2-3 times worse constraints on \fnl\ for SO compared to CMB-S4.

\section{Results}
\label{sec:results}

\begin{figure*}[t!]
\centering
\includegraphics[width=1.25\columnwidth]{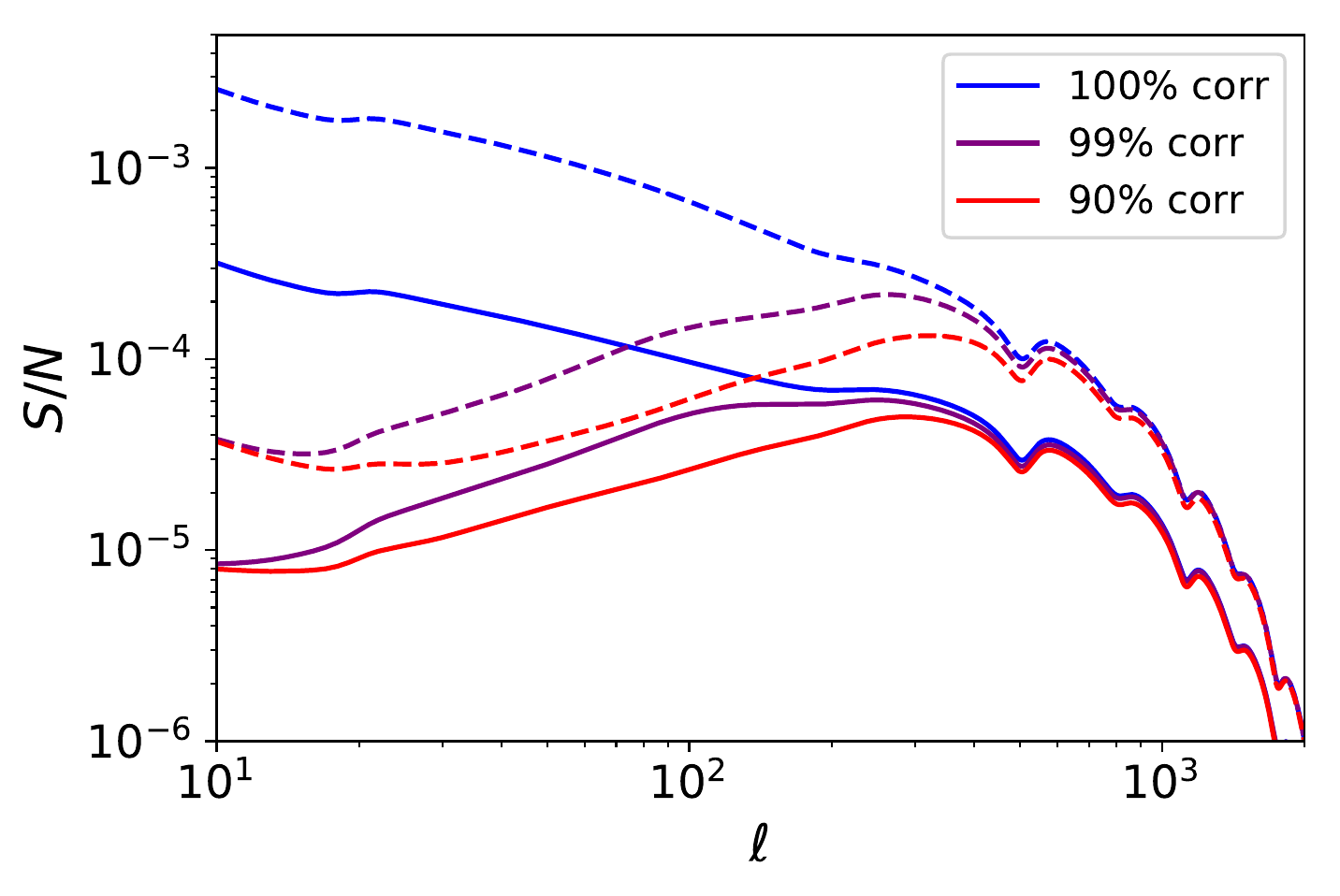}
\caption{Signal-to-noise (S/N) per multipole for various assumed values of atmospheric correlation between bands. Dashed lines correspond to S/N when only including instrumental and atmospheric noise, while solid lines also include the effects of foregrounds.
\label{fig:noise}}
\end{figure*}

\begin{figure*}[t!]
\centering
\includegraphics[width=1.25\columnwidth]{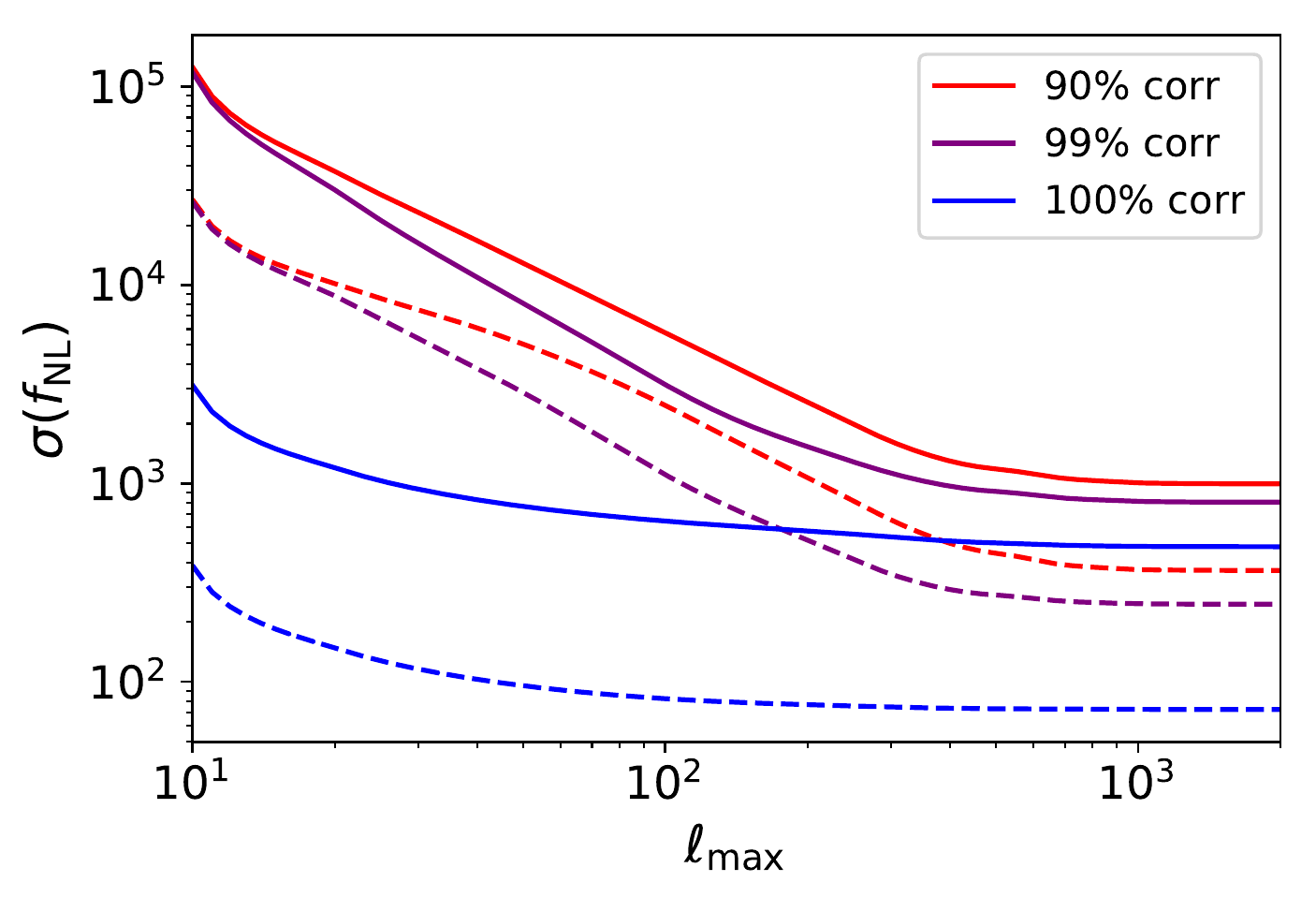}
\caption{1-$\sigma$ constraints on $\fnl$ as a function of the maximum multipole used in analysis, \lmax. Dashed lines correspond to $\sigma(\fnl)$ when only including instrumental and atmospheric noise, while solid lines also include the effects of foregrounds. In all cases, the $y$ SED has been projected out (see text for details).
\label{fig:fnl}}
\end{figure*}

Using the noise, atmosphere, and foreground paramaterizations described in the previous sections, we forecast $\sigma(\fnl)$ from the combination of $\mu \times T$ and $\mu \times E$ for the CMB-S4 experiment, combined with data from the \planck satellite \cite{Planck:2018nkj}. For our results, we assume a fiducial value of $\fnl=1$. We will generally show the cumulative constraint on \fnl\ as a function of \lmax, the highest multipole value considered in the calculation. In all cases, we assume $\lmin=10$, though our final results are not sensitive to this exact choice.

We first show, in  Figure \ref{fig:noise}, the signal-to-noise (S/N) as a function of multipole $\ell$. The solid lines include detector and atmospheric noise and foreground residual, with color differentiating the amount of assumed atmospheric correlation between frequency bands. The dashed lines correspond to detector and atmospheric noise only, with no foreground residuals. 

For the scenario where atmosphere is perfectly correlated between frequency bands 
we see that most of our signal is at the lowest $\ell$ modes. 
The addition of foregrounds reduces our overall signal by about an order of magnitude, agreeing with previous results \cite{Remazeilles:2018kqd} that foregrounds are a primary obstacle for experiments constraining $\mu$ distortions. 

The main foregrounds that impact our constraints are from Galactic sources, such as dust, synchrotron radiation, and AME. Galactic foregrounds have a larger effect on reducing sensitivities to $\mu$, compared to extragalactic sources \cite{Remazeilles:2021adt}. Even though several Galactic and extragalactic foregrounds are brighter at lower frequencies, the shape of the $\mu$ SED becomes more distinct for these frequency bands. This suggests for optimal detection of $\mu$ distortions, future experiments should concentrate on improving observations with lower frequency bands rather than higher frequency ones. In spite of foreground residuals, in the limit of 100\% atmospheric correlation between bands, we find that the lowest $\ell$ modes provide the best leverage for constraining $\fnl$.

However, when we introduce any atmospheric decorrelation, residual atmosphere has a more significant impact on our $\mu$ maps than residual foregrounds. While foregrounds reduce the S/N across the entire $\ell$ range, atmosphere suppresses the S/N primarily at low $\ell$, where the raw S/N is highest. 
This means that we lose significant leverage in constraining $\fnl$ from atmospheric contamination. 

Another potential contamination to measurements of $\mu \times T$ correlations are $y$-type spectral distortions. 
In our default component-separation algorithm, $y$ distortions are treated as noise and not explicitly projected out of the $\mu$ or $T$ map. 
However, the authors of Ref.~\cite{Ravenni:2017lgw} showed that there are $y \times T$  correlations induced from late Integrated Sachs-Wolfe effects. This means that $y \times T$ correlations will leak into our $\mu \times T$ correlations and potentially bias $\sigma(\fnl)$, unless we deproject them explicitly. This is easily accomplished in the algorithm by adding the $y$-distortion SED as one of the components in the matrix A in Eq.~(\ref{eqn:lc}), but we pay some noise penalty for this.

Figure \ref{fig:fnl} shows our constraints on $\fnl$ (including projecting out a $y$ component) after summing over all multipoles up to \lmax. We consider different values of atmospheric correlation and the presence of foregrounds.  We see that at $\lmax=2000$, we have effectively saturated our constraints on $\fnl$ for all configurations. Therefore, in Tables \ref{tab:results} and \ref{tab:deproj}, we report $\sigma(\fnl)$ for  $\lmax=2000$. 

In the ideal case in which we can ignore atmosphere, foregrounds, and $y$ distortions, we find $\sigma(\fnl) = 48$. The addition of foregrounds worsens our constraints to $\sigma(\fnl) = 475$, or $477$ if we project out $y$.  This is better than forecasts for \litebird, which is forecasted to achieve $\sigma(\fnl) = 825$ \cite{Remazeilles:2021adt}. Adding atmosphere and assuming some level of decorrelation across frequency bands, we find our constraints on \fnl\ noticeably degrade even in the absence of foregrounds. For 1\% decorrelation ($\eta = 99\%$) we find $\sigma(\fnl) = 176$, or $246$ with projecting out $y$.  The addition of foregrounds further reduces constraints, with $\sigma(\fnl) = 793$ or $804$ in the case of 1\% atmospheric decorrelation, comparable to the forecast for \litebird. We note that, because of the effect of atmosphere on CMB-S4 constraints and the low angular resolution of \litebird, it is likely that the \fnl\ constraints from the two experiments will come primarily from independent regions of $\ell$ space, in which case we will be able to improve the individual constraints by nearly a factor of $\sqrt{2}$ by combining them.

\begin{table*}[ht]
\def\arraystretch{1.5}
\setlength{\tabcolsep}{7pt}
\centering
\begin{tabular}{|l|cc|}
\hline $\sigma(\fnl)$, & \textrm{S4-Deep + \textit{Planck}} & \textrm{S4-Deep + \textit{Planck}} \\
$\lmax=2000$ & \textrm{(no foregrounds)} & \\
\hline \text{No atmosphere} & 48 &  475  \\
\hline 100\% \text{correlated} & 55 & 478    \\
\hline 99\% \text{correlated} & 176 & 793    \\
\hline 90\% \text{correlated} & 258 & 963    \\
\hline 70\% \text{correlated} & 324 &  1107    \\
\hline
\end{tabular}
\caption{
Constraints on $\fnl$ from the CMB-S4 ultra-deep survey and \textit{Planck}, considering the effects of atmosphere and foregrounds. In the case where neither are present, $\sigma(\fnl) = 48$. Including atmosphere and foregrounds worsens this constraint by at least an order of magnitude.
}
\label{tab:results}
\end{table*}

\begin{table*}[ht]
\def\arraystretch{1.5}
\setlength{\tabcolsep}{7pt}
\centering
\begin{tabular}{|l|cc|}
\hline $\sigma(\fnl)$, \textrm{``null-y"}& \textrm{S4-Deep + \textit{Planck}} & \textrm{S4-Deep + \textit{Planck}} \\
$\lmax=2000$ & \textrm{(no foregrounds)} & \\
\hline \text{No atmosphere} & 68 &  477  \\
\hline 100\% \text{correlated} & 72 & 479    \\
\hline 99\% \text{correlated} & 246 & 804    \\
\hline 90\% \text{correlated} & 364 & 995    \\
\hline 70\% \text{correlated} & 458 & 1162    \\
\hline
\end{tabular}
\caption{
Constraints on $\fnl$ when projecting out the $y$ SED, from the CMB-S4 ultra-deep survey and \textit{Planck}. We see in the case of no foregrounds that constraints on \fnl\ are worse than the ones given in Table \ref{tab:results}. Including foregrounds, we see that constraints are very similar to those obtained if we did not project out the $y$ SED.
}
\label{tab:deproj}
\end{table*}

Foreground residuals can in principle be reduced through expanded frequency coverage. If the foreground SEDs are smooth and require minimal degrees of freedom to model, with enough observations across unique frequency bands, one can constrain the foreground and CMB SEDs. Pairing CMB-S4 with additional surveys that aim to accurately model CMB foregrounds can greatly improve our constraints on $\fnl$. However, even with reduced foreground residuals, CMB-S4 will be limited by atmospheric noise to $\sigma(\fnl) >100$, unless the correlation between bands is $>99\%$.

\section{Discussion}
\label{sec:discussion}

In this work, we have presented forecasted constraints on $\fnl$, at effective scales of $k \simeq 740 \,  \text{Mpc}^{-1}$, with CMB-S4 using correlations of $\mu$-distortion anisotropies with CMB temperature and $E$-mode polarization. We find that with this ground-based experiment we are able to achieve comparable results to forecasts for the \litebird\ satellite  \cite{Remazeilles:2021adt}, depending on the amount of atmospheric correlation among frequency bands. CMB-S4 will not have prior knowledge of the amount of atmospheric correlation, which is expected to depend on the optical design and observing strategy. Co-located detectors will likely be more correlated, since they will observe along the same incident angle and will observe similar parts of the atmosphere. Conversely, detectors located further from each other will observe through slightly different columns of the atmosphere and are expected to be less correlated. Understanding the correlation properties of atmospheric emission between frequency bands is therefore vital for producing realistic forecasts of $\sigma(\fnl)$ from $\mu$-distortion anistropies.

Independent of the details of atmospheric correlation, we have demonstrated that CMB-S4 has the potential to constrain small-scale Gaussianities down to $\sigma(\fnl) \lesssim 1000$ with $\mu \times T $ and $\mu \times E$. We note that correlations between $\mu$-distortion and primary CMB anisotropies represent one of our only probes for understanding non-Gaussianities on extremely small scales or, equivalently, the ultra-squeezed-limit of the bispectrum. This presents an opportunity to constrain inflationary models that predict specific behavior at very small scales.
One such area that can be constrained with small-scale non-Gaussianities are inflationary models that contain additional fields. Certain curvaton models can naturally generate $\fnl \gg 1$ on small scales while preserving current upper limits on $\fnl$(0.05 Mpc$^{-1}$) from \textit{Planck} \cite{Dimastrogiovanni:2016aul}. 
If massive fields have direct couplings to the inflaton, they can also induce oscillatory features in the squeezed-limit bispectrum at small scales \cite{Arkani-Hamed:2015bza}. Modified (non-Bunch-Davies) initial vacuum states can also produce enhanced small-scale non-Gaussianity \cite{Ganc:2012ae} and can be constrained with  $\mu \times T$ and $\mu \times E$ correlations. Finally, $\mu \times T$ and $\mu \times E$ correlations can place stringent constraints on the running of the spectral index $n_\mathrm{s}$ and general scale-dependent non-Gaussianities \cite{Emami:2018ssb}, such as have been proposed to explain recent discoveries from the \textit{James Webb Space Telescope} \cite{Biagetti:2022ode}. 

It is also worth noting that the constraint on \fnl\ from $\mu \times T$ and $\mu \times E$ is linearly proportional to the mean value of $\mu$ (cf. Eqs. \ref{eqn:fisher} and \ref{eqn:clmux}), so any non-standard model that boosts $\langle \mu \rangle$ will correspondingly improve the forecasted constraints on \fnl \citep{Chluba:2016aln}. One example of such a model is given in  Ref.~\cite{Ozsoy:2021qrg}, in which the authors used $\mu \times T$ to constrain primordial black hole (PBH) models. PBH models generically have a rising slope for the power spectrum at small scales, preceded by a region with lower power. For scales close to the ones $\mu$ distortions. probe, the rising power can enhance $\langle \mu \rangle$, while the dip can generate non-Gaussianities larger than slow-roll predictions. These combined effects mean that upcoming CMB experiments can potentially rule out some PBH models. Work in Ref.~\cite{Zegeye:2021yml} has shown that the main signal PBH models imprint on $\mu \times T$ is from local modulation of acoustic dissipation by long wavelength modes, which induce a bias in $\mu$ distortions. We will examine this potential signal in future work.

Our analysis probes non-Gaussianities from scalar perturbations $\langle \zeta \zeta \zeta \rangle$. 
Distortions of the $\mu$ type should also be generated by primordial gravitational waves (GW) injecting energy into the photon-baryon fluid. 
This allows $\mu$ to probe the primordial tensor power spectrum. 
Since gravitational waves are free streaming throughout the radiation-dominated era, tensor $\mu^{(t)}$-distortions probe scales in the range $1 \, \textrm{Mpc}^{-1} \lesssim k \lesssim 10^6 \, \textrm{Mpc}^{-1}$, complementing the gap between CMB scales and the scales probed by upcoming GW interferometers \cite{Kite:2020uix}. 
Recent work in Ref.~\cite{Orlando:2021nkv} has demonstrated that $\mu \times B$ correlations can probe tensor and mixed tensor-scalar non-Gaussianities. Using FIRAS data, Ref. \cite{Bianchini:2022dqh} attempted the first measurement of $\mu \times B$, which they found was consistent with zero at existing noise levels.

The forecasts presented here are not the last word on \fnl\ from $\mu \times T$ and $\mu \times E$. As shown in Ref.~\cite{Cabass:2018jgj}, a cosmic-variance limited experiment can in principle constrain $\sigma(\fnl)$ to $\lesssim 10^{-3}$ with $\mu \times T$, although this will also require a significantly improved limit on the average $\mu$-distortion amplitude to break the degeneracy between $\fnl$ and $\langle \mu\rangle$   \citep{Chluba:2016aln} and ensure that no other source of distortion anisotropies is present \cite{Chluba:2022efq}. 
Putting this in context, the best constraints a cosmic-variance limited CMB experiment could place on squeezed-limit non-Gaussianity with $\langle T T T \rangle$ is $\sigma(\fnl) \sim 1$. 
One promising avenue for improvement on the constraints presented here is suggested in Ref.~\cite{Remazeilles:2018kqd}, namely expanding coverage to lower frequency ranges
where the relative amplitude of the $\mu$-distortion SED is higher.
In particular, the addition of bands at and below 10\,GHz would significantly improve detection. 
One possibility is to combine CMB-S4 with upcoming low-frequency radio surveys such as the Square Kilometer Array (SKA) \cite{Weltman:2018zrl} to improve constraints on \fnl, an aspect we will explore in future work.

\acknowledgments 
%We would like to thank.....[Wayne Gil, Peter, Dan Grin, Tristan Smith, Eric Linder]. 
T.C.~and D.Z.~acknowledge support from NSF award OPP-1852617. D.Z.~was additionally supported by the National Science Foundation (NSF) Graduate Research Fellowship Program under Grant No. DGE1746045 and the University of Chicago McCormick Fellowship. 
J.C.~was supported by the Royal Society as a Royal Society University Research Fellow at the University of Manchester, UK (No.~URF/R/191023) and by the ERC Consolidator Grant {\it CMBSPEC} (No.~725456).
G.F.~acknowledges the support of the European Research Council under the Marie Sk\l{}odowska Curie Actions through the Individual Global Fellowship No.~892401 PiCOGAMBAS.
D.G.~acknowledges support in part by NASA ATP Grant No.~17-ATP17-0162, NSF PHY-2112846, and the provost’s office of Haverford College.
V.G.~acknowledges the support
from NASA through the Astrophysics Theory Program, Award Number
21-ATP21-0135 and from the National Science Foundation under Grant No.
PHY-2013951.
J.C.H.~acknowledges support from NSF grant AST-2108536, NASA grant 21-ATP21-0129, DOE grant DE-SC00233966, the Sloan Foundation, and the Simons Foundation.
P.D.M.~and G.O.~acknowledge support from the Netherlands organization for scientific research (NWO) VIDI grant (dossier 639.042.730).
M.R.~thanks the Spanish Agencia Estatal de Investigacion (AEI, MICIU) for the financial support provided under the project with reference PID2019-110610RB-C21.

\bibliographystyle{plain}
\bibliography{main.bib}

\end{document}